\documentclass[longauth]{aa} 
\usepackage{graphicx}
\usepackage{txfonts}
\usepackage{natbib}
\begin{document}
   \title{Properties and  environment of Radio Emitting Galaxies \\ 
in the VLA-zCOSMOS survey
\thanks{based on data obtained with the European Southern Observatory 
Very Large Telescope, Paranal, Chile, program 175.A-0839 and VLA-Large Program VLA-AS0801}
}
\author{S. Bardelli \inst{1}
\and
E. Schinnerer \inst{2}
\and
V. Smol\v cic \inst{3}
\and
G. Zamorani \inst{1}
\and
E. Zucca \inst{1}
\and
M.Mignoli\inst{1}
\and
C.Halliday\inst{4}
\and
K.Kova\v{c}\inst{5}
\and
P.Ciliegi \inst{1}
\and
K.Caputi\inst{5}
\and
A.M. Koekemoer \inst{6}
\and
A.Bongiorno\inst{7}
\and
M.Bondi  \inst{8}
\and
M.Bolzonella\inst{1}
\and 
D.Vergani\inst{1}
\and
L.Pozzetti\inst{1}
\and       
C.M.Carollo\inst{5}
\and
T.Contini\inst{9}
\and
J.-P.Kneib\inst{10}
\and
O.Le~F\`{e}vre\inst{10}
\and
S.Lilly\inst{5}
\and
V.Mainieri\inst{11}
\and
A.Renzini\inst{12}
\and
M.Scodeggio\inst{13}
\and  
G.Coppa\inst{1}
\and
O.Cucciati\inst{10}
\and
S.de~la~Torre\inst{14}
\and
L.de~Ravel\inst{10}
\and
P.Franzetti\inst{13}
\and
B.Garilli\inst{13}
\and
A.Iovino\inst{14}
\and
P.Kampczyk\inst{5}
\and
C.Knobel\inst{5}
\and
F.Lamareille\inst{9}
\and
J.-F.Le~Borgne\inst{9}
\and
V.Le~Brun\inst{10}
\and
C.Maier\inst{5}
\and
R.Pell\`o\inst{9}
\and
Y.Peng\inst{5}
\and
E.Perez-Montero\inst{9,27}
\and
E.Ricciardelli\inst{12}
\and
J.D.Silverman\inst{5}
\and
M.Tanaka\inst{11}
\and
L.Tasca\inst{13}
\and 
L.Tresse\inst{10}
\and  
U.Abbas\inst{15}
\and
D.Bottini\inst{13}
\and
A.Cappi\inst{1}
\and
P.Cassata\inst{16}
\and
A.Cimatti\inst{17}
\and 
L.Guzzo\inst{14}
\and 
A.Leauthaud\inst{18}
\and 
D.Maccagni\inst{13}
\and 
C.Marinoni\inst{19}
\and 
H.J.McCracken\inst{20}
\and 
P.Memeo\inst{13}
\and 
B.Meneux\inst{7,28}
\and 
P.Oesch\inst{5}
\and 
C.Porciani\inst{21}
\and 
R.Scaramella\inst{22}
\and  
P.Capak\inst{23}
\and
D.Sanders\inst{24}
\and
N.Scoville\inst{3}
\and 
Y.Taniguchi\inst{25}
\and 
K.Jahnke \inst{26}
}

\offprints{S.Bardelli sandro.bardelli@oabo.inaf.it}  
\institute{
 1) INAF-Osservatorio Astronomico di Bologna - Via Ranzani,1, I-40127, Bologna, Italy
\and
2) Max Planck Institut f\"ur Astronomie, K\"onigstuhl 17, Heidelberg, D-69117, Germany
\and
3) California Institute of Technology, MC 105-24, 1200 East California Boulevard, Pasadena, CA 91125, USA
\and
4) INAF-Osservatorio Astronomico di Arcetri, Largo E.Fermi 5, I-50125, Firenze, Italy
\and
5) Institute of Astronomy, Swiss Federal Institute of Technology (ETH H\"onggerberg), CH-8093, Z\"urich, Switzerland
\and
6) Space Telescope Science Institute, 3700 San Martin Drive, Baltimore, MD 21218, USA
\and
7)Max-Planck-Institut f\"ur extraterrestrische Physik, D-84571 Garching, Germany
\and 
8)INAF-Istituto di Radioastronomia, Via Gobetti 101, I-40129 Bologna, Italy  
\and
9)Laboratoire d'Astrophysique de Toulouse-Tarbes, Universit\'{e} de Toulouse, CNRS, 14 avenue Edouard Belin, F-31400 Toulouse, France
\and
10)Laboratoire d'Astrophysique de Marseille,Universite' d'Aix-Marseille, CNRS , 38 rue Frederic Joliot-Curie, F-13388, Marseille, France
\and 
11)European Southern Observatory, Karl-Schwarzschild-Strasse 2, Garching, D-85748, Germany
\and
12)Dipartimento di Astronomia, Universita' di Padova, Vicolo Osservatorio 5, I-35122, Padova, Italy
\and
13)INAF - IASF Milano, via Bassini 15, I-20133, Milan, Italy
\and
14)INAF Osservatorio Astronomico di Brera, via Brera 28, I-20121, Milan, Italy
\and
15) INAF - Osservatorio Astronomico di Torino, strada Osservatorio 20, 10025 Pino Torinese, Italy
\and
16) Departement of Astronomy, University of Massachussetts, 710 North Pleasent Street, Amherst, MA 01003, USA 
\and 
17)Dipartimento di Astronomia, Universit‡ di Bologna, via Ranzani 1, I-40127 Bologna, Italy
\and
18)LBNL \& BCCP, University of California, Berkeley, CA, 94720, USA
\and
19)Centre de Physique Theorique, Marseille, France
\and 
20)Institut d'Astrophysique de Paris, UMR 7095 CNRS, UniversitŽ Pierre et Marie Curie, 98 bis Boulevard Arago, F-75014 Paris, France
\and
21)Argelander Institut fŸr Astronomie, Auf dem HŸgel 71, D-53121 Bonn, Germany
\and 
22)  INAF, Osservatorio di Roma, via di Frascati 33, I-00040Monteporzio Catone (RM), Italy
\and
23)Observatories of the Carnegie Institute of Washington, Santa Barbara Street, Pasadena, CA 91101, USA
\and
24)Institute for Astronomy, University of Hawaii, 2680 Woodlawn Drive, HI 96822, USA
\and
25)Research Center for Space and Cosmic Evolution, Ehime University, Bunkyo-cho 2-5, Matsuyama 790-8577, Japan
\and 
26) Max-Planck-Institut  f\"ur Astrophysics, D-84571 Garching, Germany
\and
27)Instituto de Astrofisica de Andalucia, CSIC, Apdo. 3004, 18080, Granada, Spain
\and 
28)Universitats-Sternwarte, Scheinerstrasse 1, D-81679 Muenchen
}

   \date{Received -- --; accepted -- --}

 
\abstract
   {}
   { We investigate the properties and the environment of 
radio sources with optical counterpart from the combined VLA-COSMOS and zCOSMOS samples. 
The advantage of this sample is the availability of
optical spectroscopic information,  high quality redshifts, and accurate density determination. }
   {By comparing the star formation rates  estimated from the optical spectral energy distribution with those based on  the radio luminosity, we divide the  radio sources in three families,  passive AGN,  non-passive AGN and  star forming galaxies. These families occupy specific regions of
the 8.0-4.5 $\mu$m infrared color- specific star formation plane, from which we extract the corresponding control samples.}
   {Only the passive AGN have a significantly different environment distribution from their control sample. The fraction of  radio-loud passive AGN increases from $\sim 2\% $ in 
underdense regions to $\sim 15 \% $ for overdensities (1+$\delta$) greater than 10. 
This trend is also present  as a function of richness of the groups hosting the radio sources. Passive AGN in overdensities tend to have higher radio 
 luminosities than those in  lower density environments. Since the black hole mass distribution is similar in both environments, 
we speculate that, for low radio luminosities, the radio emission is controlled (through fuel disponibility or 
confinement of radio jet by local gas pressure) by the interstellar medium of the host galaxy,
while in other cases it is determined by the structure (group or cluster) in which the galaxy resides.   }
   {}

\authorrunning{Bardelli et al.}
\titlerunning{Properties and environment of radiogalaxies}
   \maketitle
%

\section{Introduction}

Radio sources  have been traditionally classified into two broad classes 
depending upon their emission mechanism: AGN-driven emission has been
supposed for early type galaxies, while a star forming origin of the
radio power has been invoked for late type galaxies \citep{Condonreview}.
However, not all optical galaxies exhibit radio emission (at least at the depth
of  current surveys) and it is important to investigate  the 
physical mechanism which triggers this emission.
 For AGN-induced activity, the difference between radio quiet and radio loud galaxies 
is naturally  connected with the availability of fuel for the central engine, the black hole mass, 
 and its emission efficiency \citep[see e.g.][and references therein]{Shabala}, 
while for star forming objects it is closely related to 
the strength of the star formation episode.

Heating by a central AGN is also thought to be important for
 the host galaxy evolution, providing the energy to stop the central cooling 
that drives gas to the central black hole in an intermittent feedback mechanism 
dubbed "radio mode" \citep[see][]{Croton,Ciotti}.  
The same intermittent mechanism is indirectly observed  as bubbles 
in the intra-cluster medium \citep{Birzan2004} or in the
restarted activity of radio sources \citep{Venturi2004,Bardelli2002}.
This heating emission is thought also to be able to suppress the star formation 
and  to be an important mechanism for creating the galaxy color bimodality \citep{Croton}.
Moreover, it is a fact that many radio AGN reside in early type galaxies \citep{LO96},
a class of objects which are preferentially found in high density environments.

It has been proposed that cluster/group mergers, galaxy-galaxy mergings
 or at least tidal disturbance caused by close encounters between galaxies  
would increase radio activity.
For early type galaxies, these phenomena could drive gas more efficently 
toward the center, while for late type galaxies, these encounters
would trigger and/or enhance star formation \citep[see e.g.][]{Bekki,Vollmer}.
The radio luminosity of the AGN also depends on the interaction 
of the jet coming from the black hole with the surrounding gas 
of the host galaxy and/or of the hosting cluster \citep[see e.g.][and references therein]{Shabala}.
For these reasons, some degree of  dependence of radio activity on the environment is expected.
Although the problem is clear from the qualitative description, controversial 
results are present in the literature and for mainly two reasons.

First of all, very few papers are primarly based on radio data. 
Usually, papers in the literature consider 
AGN selected on the basis of  their optical (mostly estimated from line diagnostics) 
or X-ray activity \citep[see][and references therein]{Kauffmann2004}.
It seems that the family of optical/X-ray AGN inhabits similar environments 
to non-active galaxies, with a preference for lower densities at higher stellar masses
 \citep{Kauffmann2004,Silverman}.
 However, \cite{Best05} claim that
the phenomena generating the  emission lines and the radio emission are
statistically independent, even if recent results indicate that radio loud 
emission line AGN favour higher densities than radio quiet ones \citep{Kauffmann08} \citep[see also 
][for a study in the IR]{Caputi09}.
Secondly, only specific environments have been considered, mainly clusters 
or groups \citep{Miller,Hill, Giacintucci}, lacking therefore the coverage of a large dynamical range 
in densities. 

The first paper that studied in a complete statistical way,
both from the radio and optical side, the
environment of radio galaxies was that of \cite{Bestenv} .
This paper uses data from  2dF  Galaxy Redshift Survey \citep{Colless}
and the NRAO VLA Sky Survey \citep{NVSS} with a well 
defined estimator of density, i.e. the projected density
of the 10$^{th}$ nearest neighbor. Moreover, groups 
(and clusters) are found with the standard friend-of-friends 
detection method. The redshift range is $0.02<z<0.1$.
The general result is that radio-selected star forming 
galaxies decrease in number with increasing overdensity, while objects
with AGN activity exhibit little dependence upon the local density,
 except at the very low densities, where the probability to find these objects
decreases significantly.
\cite{Bestenv} claims also that the larger scale (i.e. on the 
scales typical of groups and clusters) is 
more important than the smaller scale (typical of galaxy pairs or companions) 
to determine the AGN radio emission.
Star forming galaxies do not exhibit dependence on environment 
at scales larger than one Mpc \citep{Kauffmann2004}.

 In this paper we investigate the environment properties of the 
radio sources present in the zCOSMOS survey using a set of well defined
density estimators in the redshift range [0.1-0.9]. 
In Section 2 we present the used data and the estimators for various
environments; in Section 3 we describe the method 
adopted to separate galaxies with radio emission induced by AGN 
from that induced by star formation and  the definition  of the control samples. 
In Section 4 we discuss the differences between the various 
samples, while in Section 5 we present the dependences on environment.
 In Section 6 the luminosity distributions  and radio-optical ratio 
for a specific class of radio AGN are 
shown and in Section 7 we discuss the results.
The adopted cosmology has $\Omega_m=0.25$,$\Omega_{\Lambda}=0.75$ and
$H_0=70$ km s$^{-1}$ Mpc$^{-1}$.

Throughout the paper we adopted the term "radio galaxy" for galaxies
hosting a radio source, independently of the physical  nature of the
 emission.

\section{Data and density estimators}

To select the optical sample, we used the zCOSMOS Bright survey, a redshift survey 
 with a magnitude selection of I$_{AB} <$ 22.5
\citep{Lilly07,Lilly09} that has been 
undertaken in the 1.7 deg$^2$ of the  COSMOS field \citep{Scoville07}, imaged
by the Hubble Space Telescope / Advanced Camera for
Surveys F814W I-band  \citep{Koekemoer}.

The observations are being carried out with the 8-meter ESO Very Large Telescope using 
VIMOS, a multi-slit spectrograph with a resolution of $R \sim 600$,  resulting in 
an r.m.s. accuracy on the velocity determinations of $\sim 110$ km s$^{-1}$ \citep{Lilly07,Lilly09}

 This paper is based on the 
first $\sim 10,000$ spectra, which have a rather non-uniform sampling pattern 
\citep[see  Figure 5 in][]{Lilly09} over an area of $1.40$ 
deg$^2$ and an average sampling rate of $\sim 33 \%$.
From this sample, we extracted a "statistical sample" of 8481 galaxies with high quality 
redshift determination ($flag>1.5$) and excluding broad line AGN.
Note that within our redshift limit of 0.9 adopted in the following analysis, 
the number of broad line AGN is 17, of which 3 are radio loud.

Absolute magnitudes were computed following the method of \cite{Ilbertmag}
from the COSMOS photometry \citep[see][]{Zuccacosmos}, which covers a wide range
of wavelength from the UV to the NIR.
 This method computes the rest-frame absolute magnitudes 
using the nearest observed band (at the redshift of the galaxy) plus a correction
derived from a template fit and is the least template dependent possible.

We used stellar masses ($M_{star}$)and star formation rates (SFRs) computed by \cite{Bolzonella} using  {\it
Hyperzmass} \citep[see details in][]{Bolzonella,Pozzetti} by means of a spectral energy distribution
 fit to the photometric data. The method uses the \cite{BC03} library and the 
\cite{Chabrier} Initial Mass Function.
The library provides a model for a simple stellar population and its evolution in 220 age steps. From this 
database   {\it Hyperzmass} imposes  a set of 10 exponentially decreasing star formation histories
 (with e-folding times ranging from 0.1 to 30 Gyrs plus a model of continuous star formation) in order to fit
the photometric data. We asssume the values of stellar mass and star formation rate 
of the best fit template.
 The typical statistical error on the stellar masses is estimated to be  $\sim 0.20$ dex.

The usually adopted conversions between radio luminosity and star formation rate
are based on the \cite{Salpeter} 
initial mass function and for this reason we corrected  our
$log M_{star}$ and $log(SFR)$ by adding
0.23 and 0.19, respectively, which are the median differences of the two quantities computed
with the two initial mass functions.

For the radio band we used the VLA-COSMOS survey, a large project consisting of 23 
pointings in an area of 1.4 x 1.4 square degrees \citep{Schinnerer,Bondi}.
 The resolution is 1.5 x 1.4 arcsec$^2$ and the survey, which has a mean r.m.s.
of $\sim 10.5$  $\mu$Jy,  detected  $2501$ sources
 at more than $5 \sigma$.

\cite{Ciliegi09} correlated this sample with the COSMOS optical samples \citep{Capak},
finding 2060 radio sources with optical counterparts.
We correlate the zCOSMOS "statistical sample" with the VLA-COSMOS sample obtaining a final radio sample
of 315 radio galaxies observed spectroscopically out of the 1231 radio 
sources with $I_{AB}<  22.5$. Note that our radio data ampling rate is 
slightly smaller than the average zCOSMOS sampling rate because the zCOSMOS survey
does not cover the entire VLA-COSMOS area.

Radio luminosities have been computed
 assuming that the radio spectrum is a power law function (defined by $S \propto \nu^{\alpha}$)
 with a slope $\alpha=-0.7$.

We used the density estimtes computed by \cite{Kovac}
 using the ZADE algorithm, which maximizes the statistical robustness of the 
estimators using both the spectroscopic and the photometric redshift 
samples.
 The algorithm offers the densities computed for each galaxy 
both as a function of the distance of the n$^{th}$ - nearest neighbour  
and in spheres of fixed comoving radius. These densities are also computed
by weighting the galaxies by either their B band luminosities or stellar masses.
As tracers of the density field, both galaxies of a flux limited and
volume limited sample have been used.

The method has been tested on mock cosmological 
simulations, by extracting samples with the same observational limits as zCOSMOS.
The overdensities were found to be correctly estimated at all redshifts
with a slight ($\sim 20\% $) underestimate of the real value at 
low densities, corresponding to $ (1+ \delta)<0.1 $ (where $\delta= \rho-<\rho>/ <\rho>$).
In the following, we will use the densities derived 
with the nearest neighbor approach. 

  In addition, the zCOSMOS optically-selected group catalog 
\citep{Knobel} offers a complementary perspective on the role 
of environment \citep{Iovino}. These groups have been extracted using a combination of the
 friends-of-friends method with a Voronoi tesselation density estimator on the
zCOSMOS spectroscopic catalogue.
The method, applied to the statistical sample, detected 
$\sim 800$ groups, 103 of which  have more than 5 members. The fraction of
galaxies in group is $\sim 20 \%$ with some dependence on  redshift.
Estimates of the group richness have been calculated using the number of galaxies of the
 volume limited sample, considered as proxies for the total richness of groups.

\section{AGN versus Star Forming galaxies in environmental studies.}

There are essentially two steps in studying environmental effects for radio sources: the division
of radio sources according to AGN  and star forming induced radio emission and the
definition of  coherent control samples.
In fact, it is already known that radio AGN  reside mostly in early type galaxies and therefore their
 overdensity distribution has to be compared only with that of radio quiet early type  galaxies, 
otherwise any possible  difference observed in environment could  just be due to the optical morphology-density relation
\citep{Dressler}.

However, a sharp division between AGN and SF galaxies in some
 cases has little physical meaning, because 
we know that both phenomena are commonplace in a number of galaxies 
  \citep[see e.g.][]{Silverman2} and
it is therefore difficult to determine the fraction
of the radio flux that originates from AGN and/or from the star formation
activity.

 For AGN/SF galaxy  luminosity function studies \citep[e.g.][]{Bardelli,Smolcicagn,Smolcicsf}, the aim 
is to count the  total number of AGN (or star forming galaxies) and for these
studies it is correct to consider 
all objects exhibiting AGN (SF) activity, regardless of its relative contribution to the radio flux. 
For density studies the situation is slightly different. 
In this case, we ask whether the environment has (or not) any effect in triggering radio emission from 
AGN or star formation and it is not easy to reach a conclusion in those composite objects, considering that
the trigger mechanisms are likely very different.

In other words, if a class of radio sources exhibits an environmental dependence, is  this related to 
its star forming or AGN part of the flux?

In the following, we present a method for selecting sub-samples in such a way that  the "radio AGN sample" contains radio sources for which 
the AGN emission is at least about one order of magnitude higher than that
expected from the star formation.
 Therefore, our definition of radio AGN is "an object for which
the radio emission is largely dominated by the emission of the central engine".

With this definition, the optical AGN as defined on the basis of emission line ratios, will be included 
in the star forming galaxies sample if they have a low ratio of AGN/SF radio flux.  
This inclusion is supported by  a) the independence of emission line and radio emission \citep{Best05},  at least for low luminosity AGN as ours, 
b) the independence of these AGN from the environment \citep{Silverman}, 
similar of what has been {\it a posteriori} found for star forming galaxies and c) the number of
optical AGN with  radio emission, which  is negligible with respect to that of star forming galaxies.
 \begin{figure}
   \centering
   \includegraphics[width=\hsize]{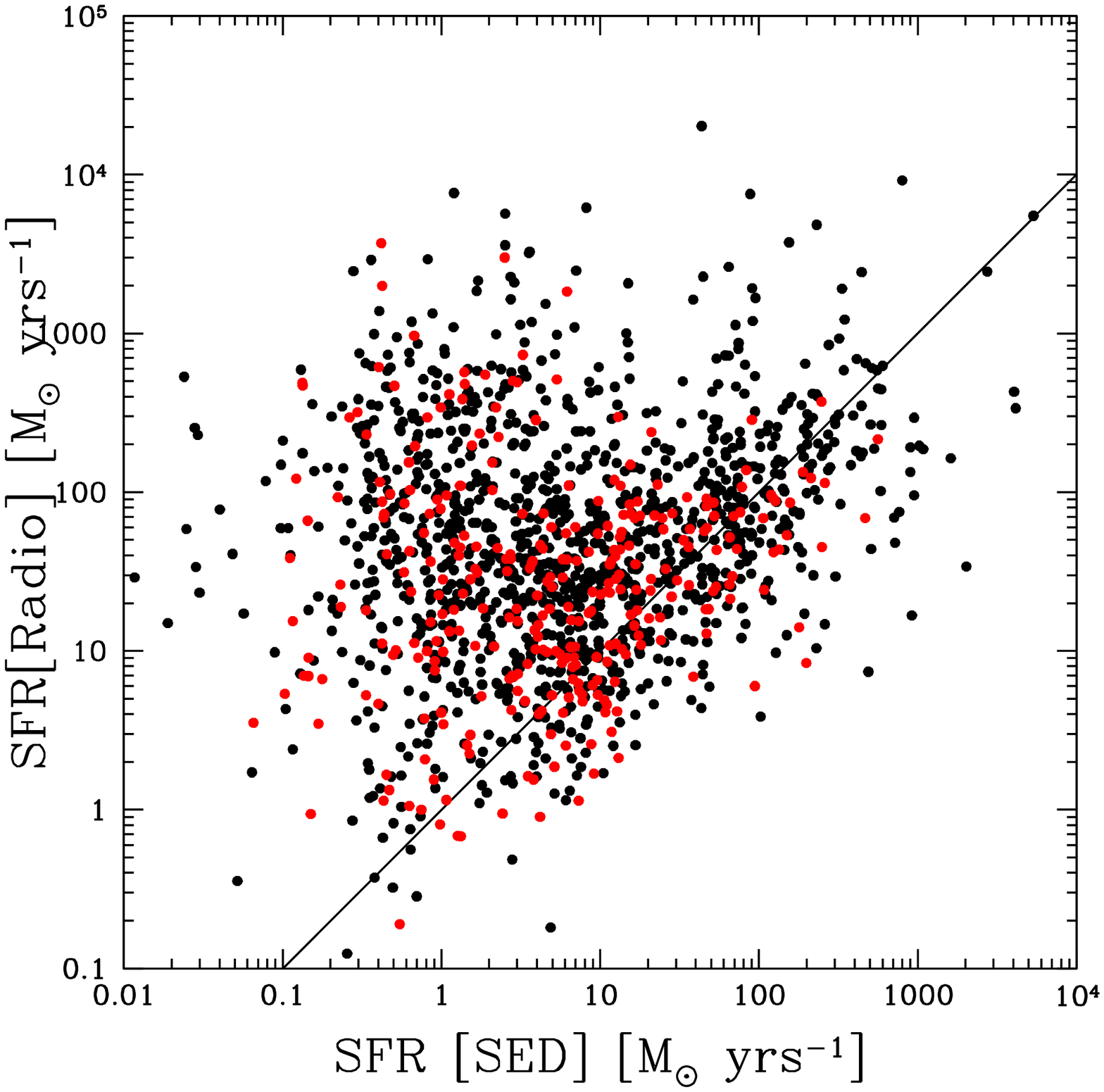}  
\includegraphics[width=\hsize]{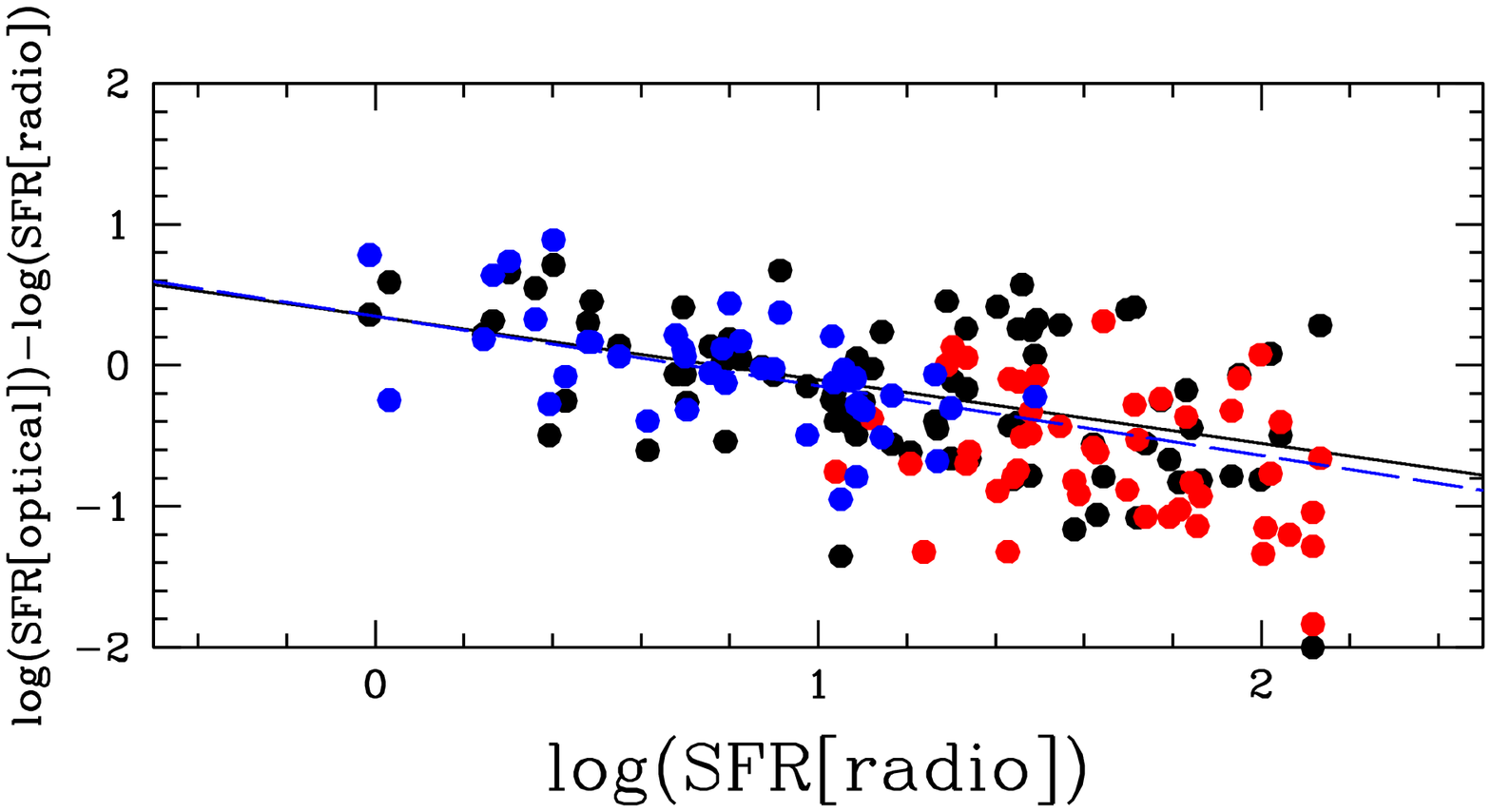}
\caption{Upper panel: SFR(radio) versus SFR(SED) for all radio galaxies in the initial 
VLA-COSMOS sample with $I<22.5$ (black circles).
 In red the 315 objects within the zCOSMOS survey are highlighted. 
The solid black line is  the SFR(SED)=SFR(radio) line drawn for reference.
Lower Panel: Ratio of the SFR as determined in the optical (from SED and lines) 
and the SFR estimated from  radio luminosity for a sample of "bona fide" star forming galaxies 
(see text).
In red the SFRs obtained from the [OII] line, while in blue from the H$\alpha$ line. Black points
refer to SFRs determined from SED. Blue and black lines are the least squares fits to line and SED
determination.}
         \label{sfrsfr}
   \end{figure}

\begin{figure}
   \centering
   \includegraphics[width=\hsize]{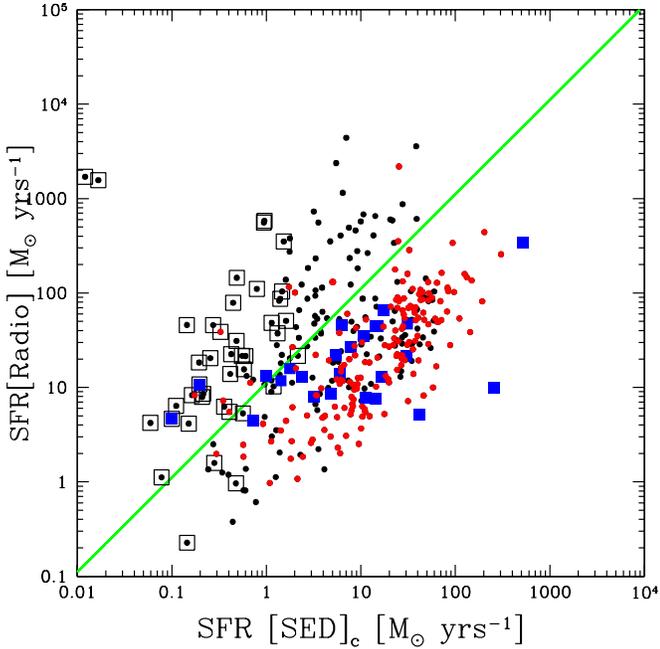}
\caption{ SFR(radio) versus the corrected (see text) SFR(optical) for our zCOSMOS sample. 
Red points are star forming galaxies, blue squares are optical
AGN  (LINERs and Seyfert 2) as determined by line diagnostics,
while   empty squares are radio emitting objects in the passive region in the infrared colors-specific star formation plane (see Section 3.2).
 The green line 
is the adopted division between star forming galaxies and AGN. Small black points are objects without spectral diagnostic classification (due to
low quality line measurements).  }
         \label{sfr_sfsy}
   \end{figure}

\subsection{AGN versus star formation radio emission}

There are several recipes for dividing the radio population into sources with emission from an AGN or 
from star formation (SF) \citep[see e.g.][and references therein]{Smolcicmethod, Best05}.
They are mainy  based on determining the optical properties of known  radio emitting star forming galaxies and 
 defining a limit for the emission of the AGN. In particular, the method of \cite{Best05} separates the
 two classes on the basis of the D$_n$(4000)-L$_{1.4GHz} /M_{star}$ plane. In practice, this method compares the ages of stars 
(estimated with the  D$_n$(4000) parameter) with the current specific star formation rate (estimated with L$_{1.4GHz} /M_{star}$) and the 
chosen separation line is 0.225 above the  D$_n$(4000) value of a 3 Gyr exponentially 
decaying star formation galaxy  track. 
The method of \cite{Smolcicmethod} is based on the fact that the two families define two 
peaks in a PCA-based rest-frame colors combination. Both methods have been calibrated with the use of line diagnostics.

Our method  instead has two steps. First, we compare the star formation rates estimated from the radio 
emission and from the spectral energy distribution  fit \citep{Bolzonella}  and we
 separate AGN from SF radio sources on the basis of this comparison.
For the estimate of the star formation rate from the radio emission we used the \cite{Bell}  formula.
 Second, we study the optical and infrared 
properties of the radio galaxies  to define  control samples with similar properties.

Note that our results do not depend strongly on the specific formula adopted in converting $L_{radio}$ to SFR  
because our method is based on a relative comparison, which considers as AGN all objects outside a
given  radio-optical star formation correlation.
With the zCOSMOS spectra  we could also use the star formation rates 
derived from the lines, but we decided to use the SED determinations 
because: a) at low redshifts ($z<0.5$) the emission line estimate would use
the H$\alpha$ line (which is affected by  problems with sky line contamination and strong fringing in the spectra  in the 
redshift range [0.3-0.5]),
while at higher redshift the usable line is [OII] and b) for weak star formations the
line measurements are able to provide only upper limits. 
 We note that because of point a), the SFR estimate would not be homogeneous being based on two different lines
in two different redshift ranges.
Moreover, by  considering only detected lines, the final usable sample would be $\sim 60 \%$ 
of that we have with SED determinations. For this reason, we will use the spectral features only
to calibrate the method.

In Figure \ref{sfrsfr} (upper panel)  we show the SFR(radio)-SFR(SED) plane for the  entire VLA-COSMOS sample limited to
$I<22.5$)
and its zCOSMOS subsample (red points). Only in this case, used for illustratory purpose,  we used the photometric redshifts as 
derived by \cite{Ilbert} both for computing distances and  SEDs.

 It is clear that at high ($>10 M_{\odot}$ yr$^{-1}$)  values of SFR(SED),
  the two quantities are reasonably correlated, defining a "star forming galaxy line", while at 
low values (SFR(SED)$<2 M_{\odot}$ yr$^{-1}$) a significant population
 of radio sources with a radio star formation "excess" is present. This excess means that the radio emission of this population  is not driven by 
star formation and thus is due to AGN. 
 
However, we need to test the robustness of the SED star formation rate determination. By taking advantage of having the zCOSMOS spectra, we can apply standard line diagnostics 
\citep{Baldwin}  to classify emission line galaxies as
star forming, Seyfert 2 or LINERS. This procedure was applied in an automated way to the
zCOSMOS sample \citep{Bongiorno} following the method of \cite{lamareille}.
 At low redshifts ($0.15<z<0.45$) the [OIII]/H$\beta$  versus [NII]/H$\alpha$ ratios have been used 
(red diagnostics), while at higher redshift ($0.5<z<0.93$) the objects have been classified using
 the line ratios [OIII]/H$\beta$ vs [OII]/H$\beta$ (blue diagnostics).

From the 133  radio detected galaxies classified as star forming, we selected a subsample of 83 "bona fide star forming galaxies",
 i.e. classified as SF galaxies from diagnostics and with high quality line measurements and with emission lines far from sky and fringing lines.
In Figure \ref{sfrsfr} (lower panel) we plot the ratios of the SED (black points)
or lines (red and blue points) SFR to the radio SFR determination versus the radio SFR.
Red points correspond to SFR obtained from the [OII] line  and  blue  points from the H$\alpha$ line
\citep{Moustakas}. Note that SED and emission line determined SFR are consistent with each other, but 
their ratios with the radio SFR shows a significant decrease with increasing SFR(radio).   

 In particular, assuming that the values of the radio SFR represent the true ones,
the SFR derived from SED and emission lines 
 overestimate the low  and underestimate the high SFR. We note that the underestimate of the SFR
from the H$\alpha$ line (red points) for $\rm{SFR}>1  M_{\odot}$ yr$^{-1}$  is qualitatively similar to the result of  \cite{Caputia},
 found by comparing this estimation with the value derived
 from the combined UV and IR luminosity
 (see their figure 8).

 We fitted  the points in the figure (the "bona fide star forming galaxies") with a
 linear relation finding that $ <log(\rm{SFR}[SED])> = 0.55 log(\rm{SFR}[Radio]) + 0.347$,
where $<log(\rm{SFR}[SED])>$ is the mean value of log(SFR[SED]) at a given 
 $log(\rm{SFR}[Radio])$. 
Then, we computed the "corrected" $\rm{SFR}[SED]_c$ by applying the relation
$$log (\rm{SFR}[SED]_c)=log (\rm{SFR}[SED])-<log (\rm{SFR}[SED])>  $$
$$+log (\rm{SFR}[Radio]) $$.
In practice, we force the two estimates of SFR for the "bona fide star forming galaxies" to statistically follow the 
relation $log( \rm{SFR}[SED]_c)=log (\rm{SFR} [Radio])$
 The  r.m.s. dispersion around the fit is $\sim  0.35$.

Finally, we define the division line between SF galaxies and AGN to be the line parallel to
the  $log (\rm{SFR}[SED]_c)=log (\rm{SFR}[Radio])$ relation 
and shifted horizontally by 3 sigma.
In this way, to be included in our AGN sample, the radio emission 
from the black hole should be about an order of
magnitude higher than the average emission due to  star formation. 

Obviously, this division line between star formation and AGN 
is somehow subjective, because it is likely that there 
is some AGN contribution to the radio flux also in objects 
 below the division line.
 However, this procedure ensures that we define a "clean", even if not complete,  AGN sample.
 In particular, at higher star formations rates, we cut progressively at higher radio luminosities.

Applying this procedure we define a sample of 97 of AGN out of our total sample of 315 radio sources
(Figure \ref{sfr_sfsy}).

Among the 133 automatically diagnosed star forming galaxies, only 13 are in the AGN region.
 After visual checks of the spectra, we conclude that all but two could be considered incorrect or at least uncertain classifications 
because of a low  signal-to-noise  of the lines or contamination by the sky/fringing lines. This shows that a large fraction ($>90 \%)$ 
of the objects classified as star forming galaxies on the basis of the optical spectral diagnostics are indeed 
recognized as such by our method.

As expected, the situation for optically classified AGN is less clear: among the 14 LINERs and 11 Seyfert 2 objects, the galaxies 
showing "star formation excess" are 2 and 3, respectively. Assuming that for
our optical AGN  the average contribution of the AGN to the line luminosity is $\sim 50 \%$  \citep[see figure 6 of][]{Silverman2}
and  comparing the expected radio star formation from the optical-radio star formation correlation
with the observed radio luminosity,  we conclude that, on average, the AGN radio luminosity for these objects is a factor
$\sim 3-4$ higher than that due to star formation. 
 
 Note that although  the SFR(optical)-SFR(radio) plot is useful for distinguishing  AGN from star forming galaxies, it can not be used 
to extract a consistent control sample, because the division  between galaxy classes is not derived  using optical properties only.

\subsection{The color-specific star formation plane}
 \begin{figure}
   \centering
   \includegraphics[width=\hsize]{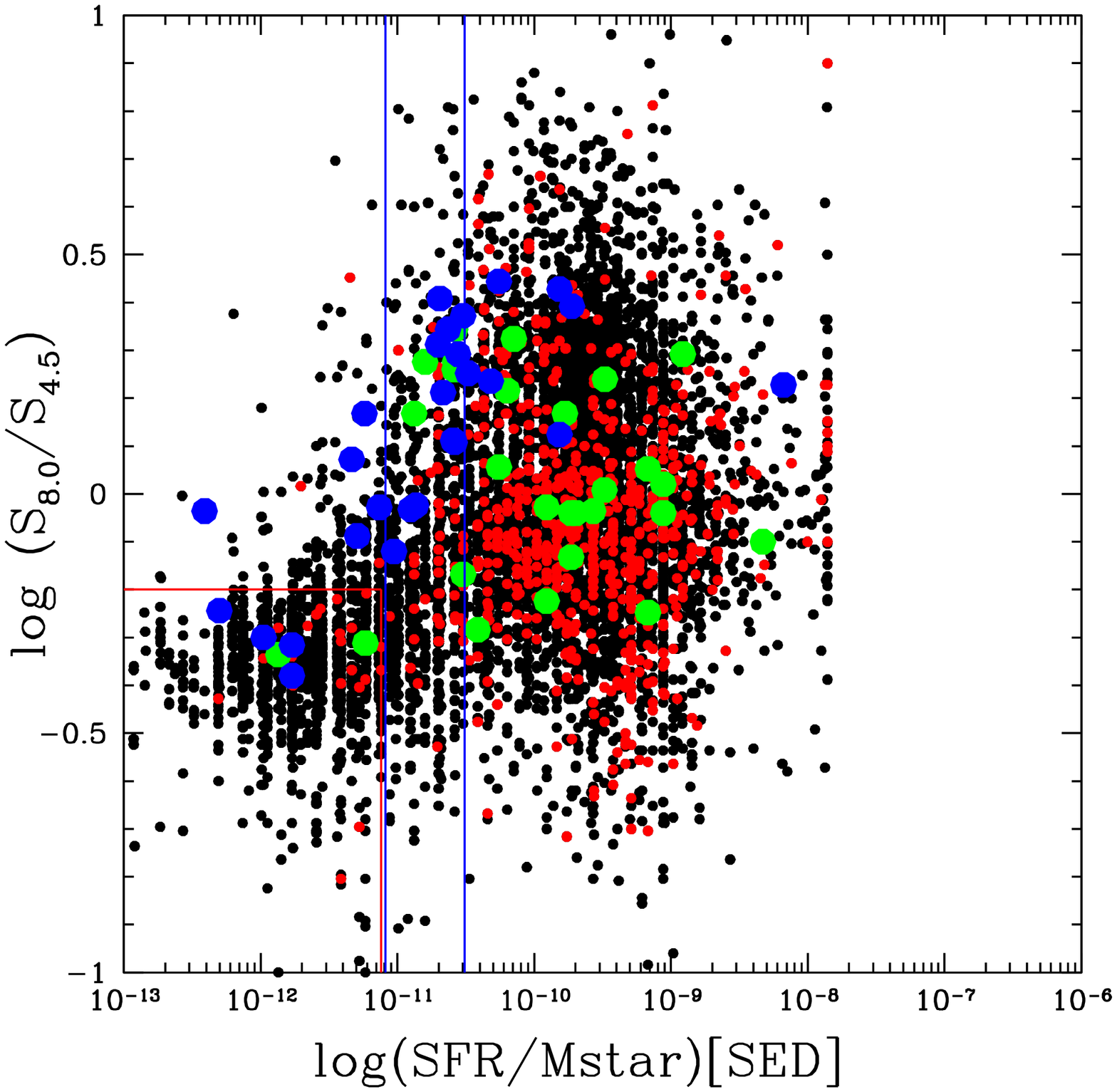}
   \includegraphics[width=\hsize]{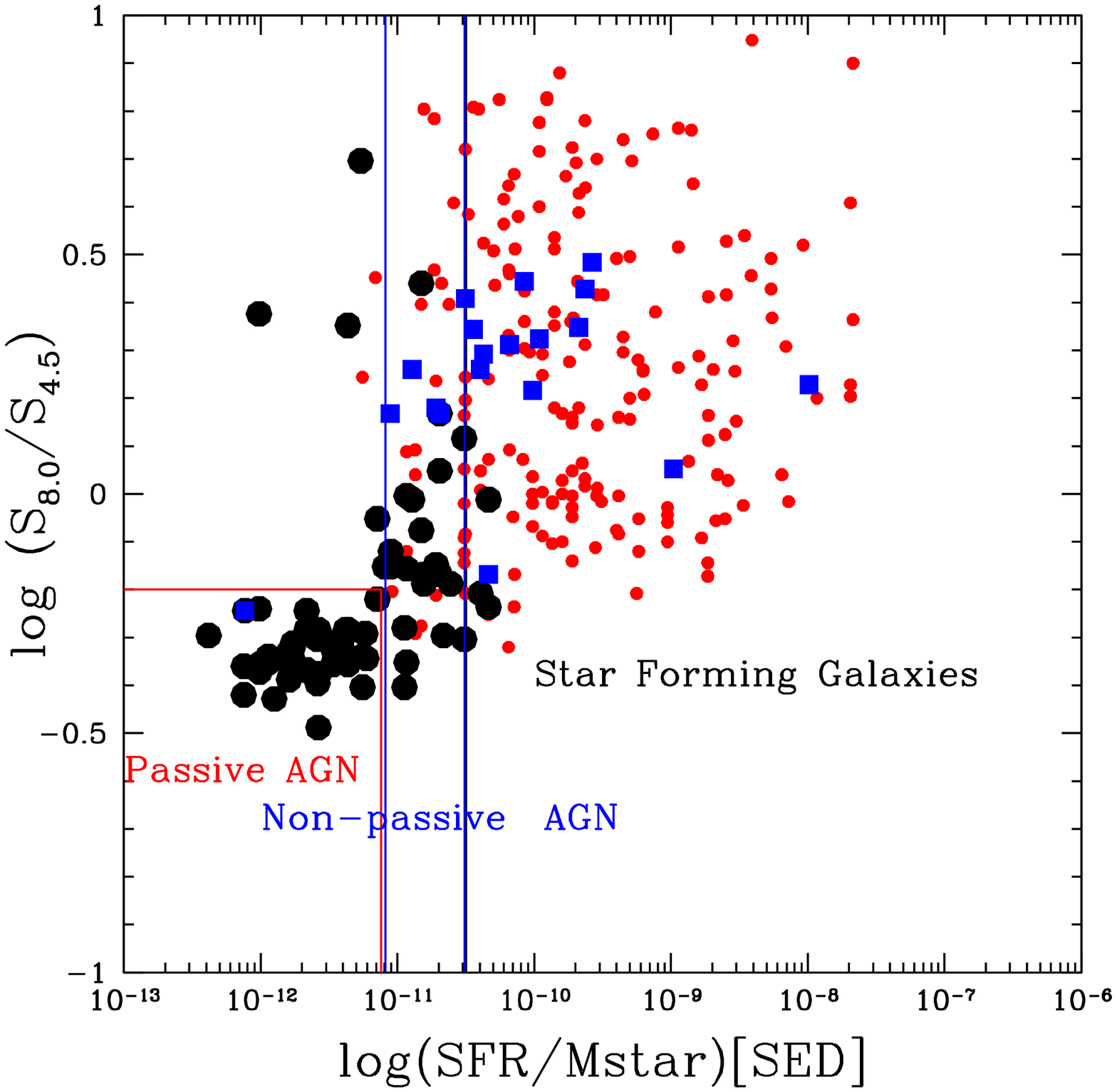}
\caption{Upper panel: infrared color versus specific star formation rate for all galaxies in 
the statistical sample. Red points represent the star forming galaxies, blue points the LINERS and green points the Seyfert 2 objects  classified by line diagnostics.
Black points represent unclassified galaxies (see text).  Lower panel: same as the upper panel, showing only the radio galaxies. The large circles  represent our defined radio AGN. Blue squares are the optical AGN (not divided into LINERs and Seyfert 2). Vertical and horizontal lines represent the division between the three
galaxy populations defined in the text.}
         \label{colmstar}
   \end{figure}

We now discuss the relation between the infrared color $8.0$-$4.5$ $\mu$m 
obtained by the Spitzer satellite \citep{Sanders}
and the optical specific star formation rate (SSFR). It is already known that near infrared colors are a powerful
method for finding broad line AGN \citep[see e.g.,][]{Lacy}, but here we are more interested in the behaviour of the
"normal" galaxy population. 
We define the infrared color as ratio of the two fluxes  
$S_{8.0}/S_{4.5}$ 

In the upper panel of Figure \ref{colmstar} we show the distribution of the galaxies of the statistical sample (as defined in
Section 2),
where the star forming galaxies  are indicated in red,  the Seyfert 2 galaxies in green and  the LINER galaxies in blue, all classified with 
spectroscopic diagnostics.
Note a number of non-classified objects (black points) that reside in the star forming galaxies, where the emission lines are expected to be strong. These
classification failures can be explained by noting that the redshift distribution of these objects has two peaks at $z\sim 0.5$ and $z\sim 0.85$.
In the first case, we are at the point in which the H$\alpha $ line 
is at the red border of the observed spectrum and affected by fringing, and the [OII] line starts to enter from the blue border. At the peak at high redshift, the H$\beta$ starts to be affected by fringing.

 The existence of two populations is rather obvious and  is related to  the bimodality found  also in the optical by using B-I or U-B colors \citep[see][]{Cucciati} 
 for dividing red, passive galaxies from the star forming ones. 

Interestingly, the Seyfert 2 objects occupy the same locus of star forming galaxies, while LINERs are at the upper border
\citep[see also figure 19 of][]{Smolcicmethod}.

Passive galaxies are located in a clump in the lower left of the plot, while 
the star forming galaxy
region has a vertex nearby the passive clump and opens at larger distance.
  The division value has been chosen as the point in SSFR which show a minimum in the density distribution of the galaxies.

In the lower panel of Figure  \ref{colmstar}, we show the same plot but only for radio sources. The large black circles are the radio  AGN objects  defined in the previous subsection. 

This plot shows that the radio AGN sample consists of two different populations, one corresponding to the passive galaxies ($log(S_{8.0}/S_{4.5})<-0.2$ and  
 $log(SFR/Mstar) < -11.1$)  and one extending toward the SF galaxies locus.

Moreover, it also appears as if there is a  specific star formation limit  (at $log(\rm{SFR}/Mstar) \sim -10.5$), such that the radio AGN are rare for SSFR higher than this limit. 
The division at $log(SFR/Mstar) \sim -10.5$ is certainly subjective and may 
depends on the
characteristics of the survey, but it is a fact that the majority of radio AGN are on the left of the line and are not 
spread throughout  the plot. This division  value has been chosen to maximize the number of radio AGN and  minimize the contamination from SF galaxies in the region defined for AGN.

Of the 75 AGN radio sources with measured infrared color, 38 are  passive galaxies, while the remainig are  
in the second region defined above.  
We note that this behaviour is similar to that shown in  figure 10 of \cite{Smolcicmethod}  using the $8.0$-$4.5$ $\mu$m and  $5.8$-$3.6$ $\mu$m color plane.

 In Figure \ref{sfr_sfsy}, there are a number of radio sources defined as passive which are not  in the passive AGN sample, because they are not radio luminous  enough
to satisfy our cut. Assuming the extreme hypothesis that all these objects are passive AGN, this would mean an incompleteness of our sample
of $\sim 15 \%$. Recomputing all quantities with the inclusion of these objects, the following results do not change.

As a general conclusion, we define as radio AGN  all objects with "radio star formation excess", while 
"passive AGN"  occupy the passive  (i.e. very low specific star formation) 
galaxies peak of  Figure \ref{colmstar}.
 The remaining AGN are defined as "non-passive
AGN", while all other radio sources are considered to be star forming galaxies.

  As outlined above, 
this choice is justified in the light of the analysis of the environmental
 dependencies. It is worth  emphasizing that
a more sophisticated method to separate AGN from starforming galaxies, when
 both phenomena are present, is still lacking in the literature. 
Our method leads to a contamination of the star forming galaxy
 sample  by AGN of $\sim 8 \% $ (after 
having applied the limits of  Table \ref{numtab}).    We checked that 
this contamination has no effect in the following conclusions.

Therefore, the control sample will be extracted using the limits taken from Figure \ref{colmstar} and reported in Table \ref{numtab}.
The only exception is the additional cut that we applied to the stellar mass in the passive AGN sample, which is discussed in Section 4.2.

\begin{table*}
\centering          
\label{table:1}     
\caption{Sample definition: numbers in the third and fourth columns are with and without the 
"evolving" magnitude limit}  
\begin{tabular}{l l l l l}     
\hline     
\smallskip
sample & limits &  Number in radio sample & Number in Control Sample \\
\hline                    
\smallskip
 "passive AGN" & log(SFR/Mstar)[SED]$<$-11.1 \& log($S_{8.0}/S_{4.5}$) $<$-0.2 &
38/38  & 524/530         \\ 
\smallskip
 "non-passive AGN"   &   -11.1$<$log(SFR/Mstar)[SED]$<$-10.5 & 31/37 & 496/581    \\
\smallskip
star forming galaxies &  log(SFR/Mstar)[SED]$>$-10.5 &  154/186  & 1672/3154 \\
\hline                  
\end{tabular}
\label{numtab}
\end{table*}
%

\subsection{Comparison with other classifications}

\begin{figure}
   \centering
\includegraphics[width=\hsize]{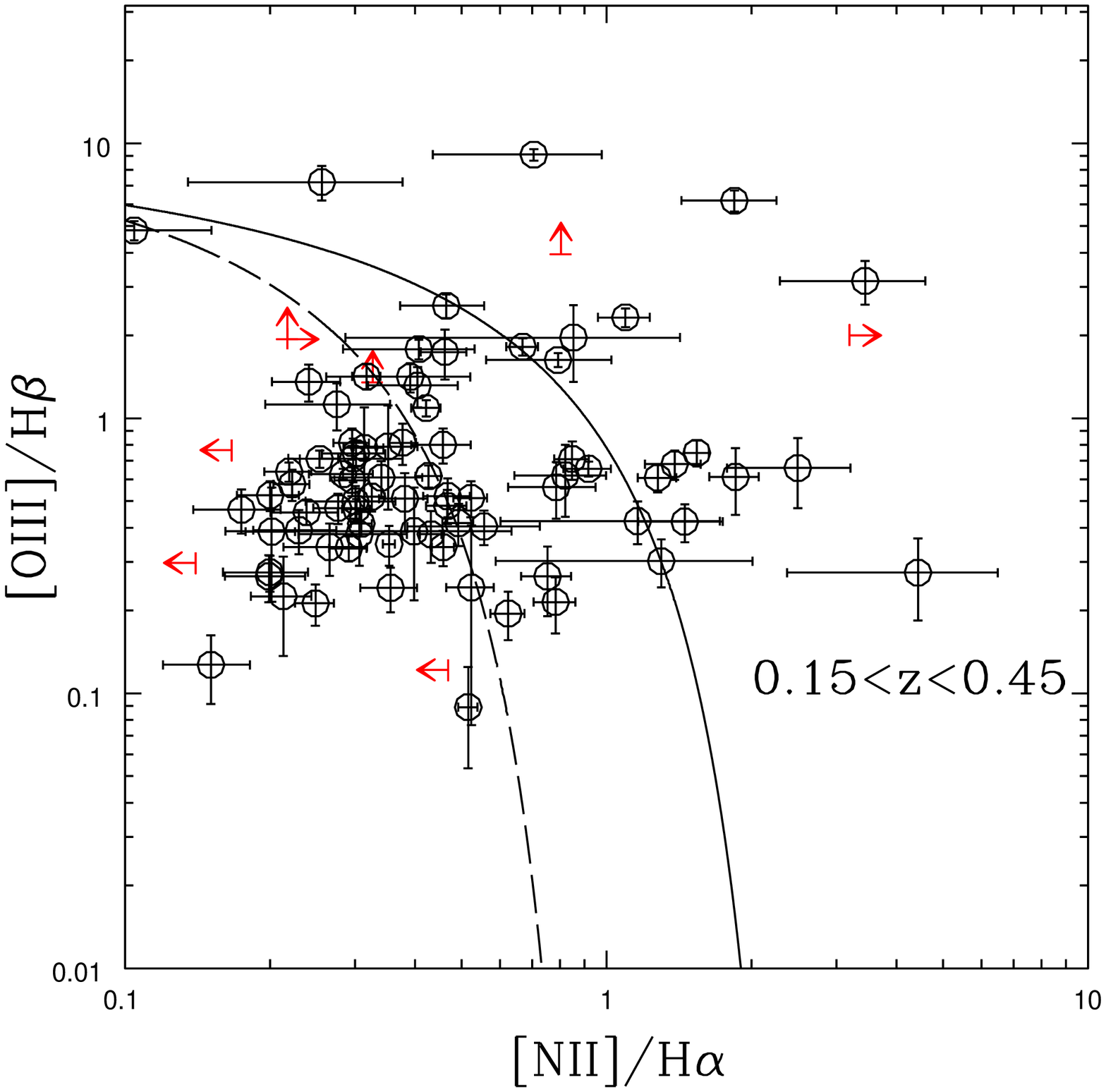}
\includegraphics[width=\hsize]{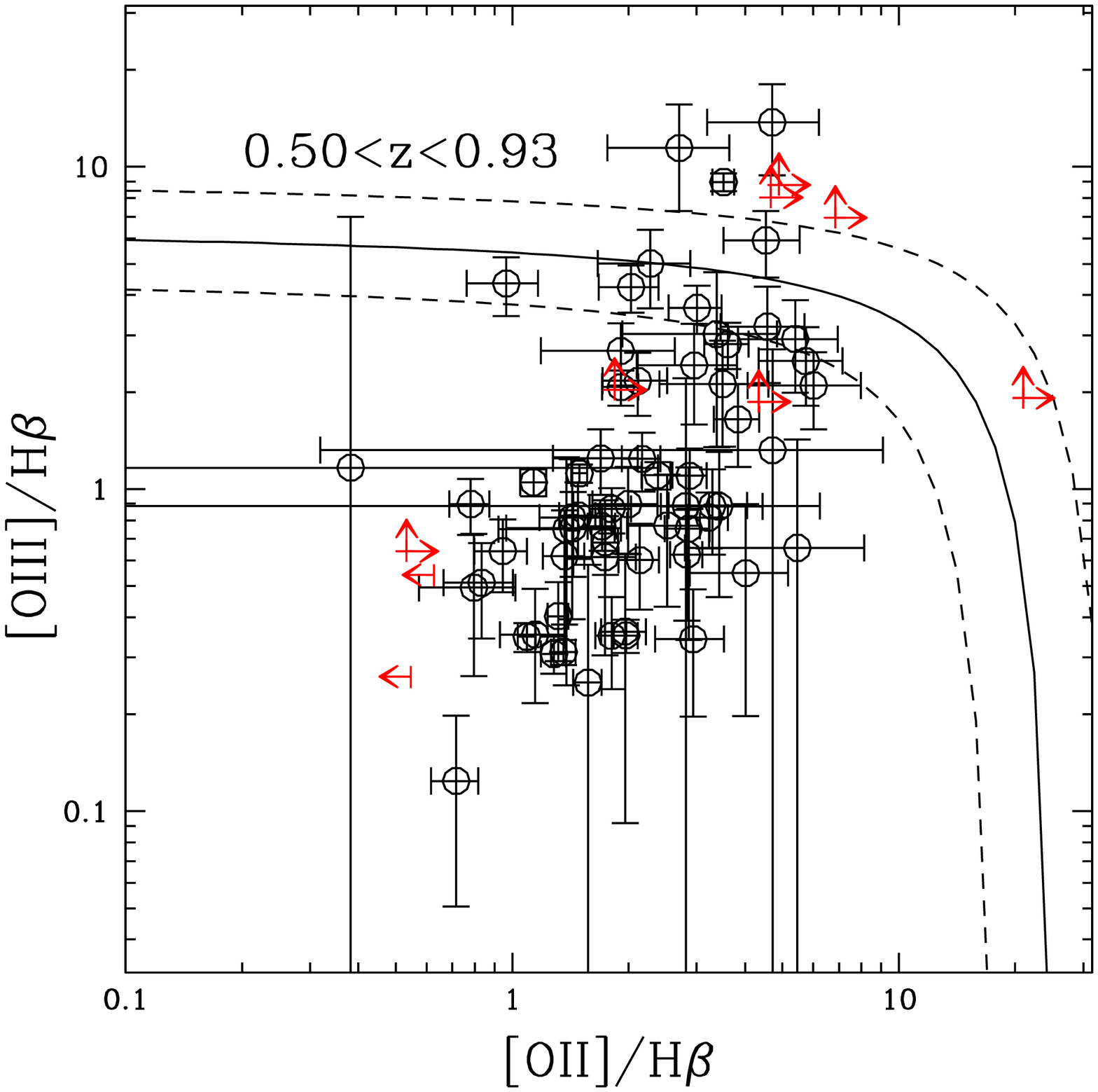}
\caption{Red and blue diagnostics diagram for our star forming galaxies with radio detection. Circles are objects with all lines
detected. For the upper or lower limits, we plot only objects with only one upper limit determination for each line ratio.  }
       \label{diaSF}
   \end{figure}

To check the consistency between  our method for classifiying AGN and SF galaxies and other methods present 
in the literature,
we will consider the spectroscopic line diagnostics \citep{Baldwin}, the  D$_n$(4000)-L$_{1.4GHz}/M_{star}$ plane of 
\cite{Best05} and the \cite{Smolcicmethod} method. 
In Figure \ref{diaSF}, we present a plot of the equivalent width of the [OIII]/H$\beta$  versus [NII]/H$\alpha$ (red diagnostics) 
and the  [OIII]/H$\beta$ versus [OII]/H$\beta$ (blue diagnostics)  line diagnostics for radio SF galaxies as defined by our method. 
We remind that the two diagnostics 
were applied to two redshift ranges ([0.15-0.45] and [0.5-0.93], respectively).
Circles are objects for which all emission lines were
detected.
 Here, an upper or lower limit implies that  for each line ratio one line has been detected and for the other we have an upper limit. When we have upper limits for both lines, no point is shown in the diagram.

In the upper panel, the solid line is the theoretical \cite{Kewley}
division between SF galaxies and AGN, while the dashed line is the \cite{Kauffmann03} line. The region between the 
\cite{Kewley} and the \cite{Kauffmann03} lines defines a region of so-called `"composite galaxies".
In the lower panel, the division proposed by \cite{Lamareille2004} is plotted as solid line and dashed lines represent the $+/- 0.15$ dex 
uncertaintes in the classification.
  
Of the 100 SF galaxies within the redshift range useful for the red diagnostics, 81 have spectra with 
all four lines detected,  while for the blue diagnostics the ratio is $63/91$.  
As expected from Figure \ref{sfr_sfsy}, a not negligible number (14) of the optical AGN (i.e.
those defined by line ratios) are in the SF galaxy sample.

For the red diagnostics, 18 objects fall in the "composite galaxy region": by applying the same procedure 
defined in Sect. 3.1 to estimate the relative 
contributions of  AGN and  star formation rate to the radio flux,  for these objects we obtained  on average $\sim 2.5$.
For the non-passive and passive AGN samples, the diagnostics are inconclusive, because of the lack of detected lines for estimating the ratios.
 The blue and red  diagnostics give a different ratio between the number of detected AGN and star forming galaxies.
 This is related to the different efficiency in detecting AGN of the two diagnostics, as noted in \cite{Bongiorno} (see their Figure  3 and 4).
In particular, the blue diagnostics classify a large number of X-ray emitting AGN as star formimg galaxies,
reinforcing therefore the need of new AGN-star forming division methods.

\begin{figure}
  \centering
   \includegraphics[width=\hsize]{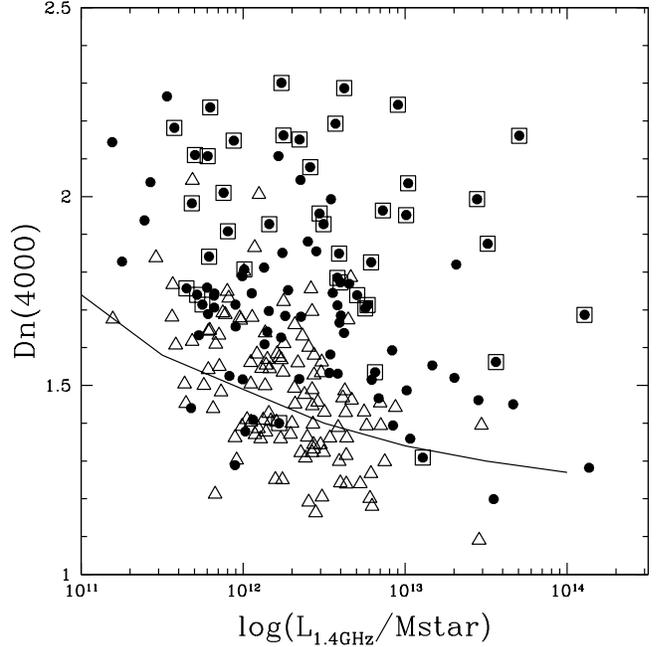}
\caption{ Dn(4000) versus L$_{1.4 GHz}$/Mstar plane: Triangles are our star forming galaxies and circles our AGN sample. Open squares indicate the passive AGN.}
  \label{bestplane}
   \end{figure}

Another method for differentiating between  AGN and SF galaxies through the use of the  D$_n$(4000)-L$_{1.4GHz}/M_{star}$ plane  \citep{Best05},
where the age of the galaxy stellar population (estimated by   D$_n$(4000)) is compared with the specific star formation rate derived from the radio luminosity. The division line adopted by \cite{Best05} is 0.225 above the  D$_n$(4000) value of the track defined by a 3 Gyr exponentially decaying star formation. In Figure \ref{bestplane} we show our samples in this plane.
The  number of AGN below the division line is 5 (out of 104), while $\sim 50 \%$ of our SF galaxies are above the \cite{Best05} dividing line and they would be classified as AGN on the basis of this criterion.
 These galaxies are likely in a transition region, where our AGN induced emission objects are mixed with the SF ones.

The \cite{Smolcicmethod} PCA based classification is another, complementary way for classifying AGN and star forming objects.
Of the 129 AGN defined in \cite{Smolcicmethod}  in common with the zCOSMOS sample, 66 are star forming excess 
galaxies in our classification scheme (and therefore AGN), while 9 out of 97  of the \cite{Smolcicmethod} star forming radio galaxies are in our AGN locus.
The fraction of AGN that have $log(SSFR)<-10.5$ is $\sim 60 \%$.
A similar percentage of star forming excess was found  in  the AGN sample as defined in \cite{Bardelli},
where a spectrophotometric division is adopted.
The apparently large difference between  our and the \cite{Smolcicmethod} method is largely  
due to the fact that optical AGN and composite galaxies are included in the AGN sample in \cite{Smolcicmethod} and not in our scheme.

In fact, by applying our diagnostics to the \cite{Smolcicmethod} sample, we confirm that the contamination of their 
AGN and star forming sample is   $\sim 17 \%$ and $\sim 10\%$, respectively consistent with what is estimated in their paper.

A general conclusion of this comparison is  that our approach, which entirely relies on the comparison of observed  radio power with the expected power due to the star formation, is more conservative (and therefore not complete) in selecting radio AGN than other methods. We remind that our method was constructed to classify as AGN those objects for which the radio flux is 
dominated by the central engine.

\section{Differences between radio loud and control samples}

\subsection{Spectral differences between radio loud and control samples}

The most basic approach is to construct composite spectra of the samples. 

\begin{figure}
   \centering
   \includegraphics[width=\hsize]{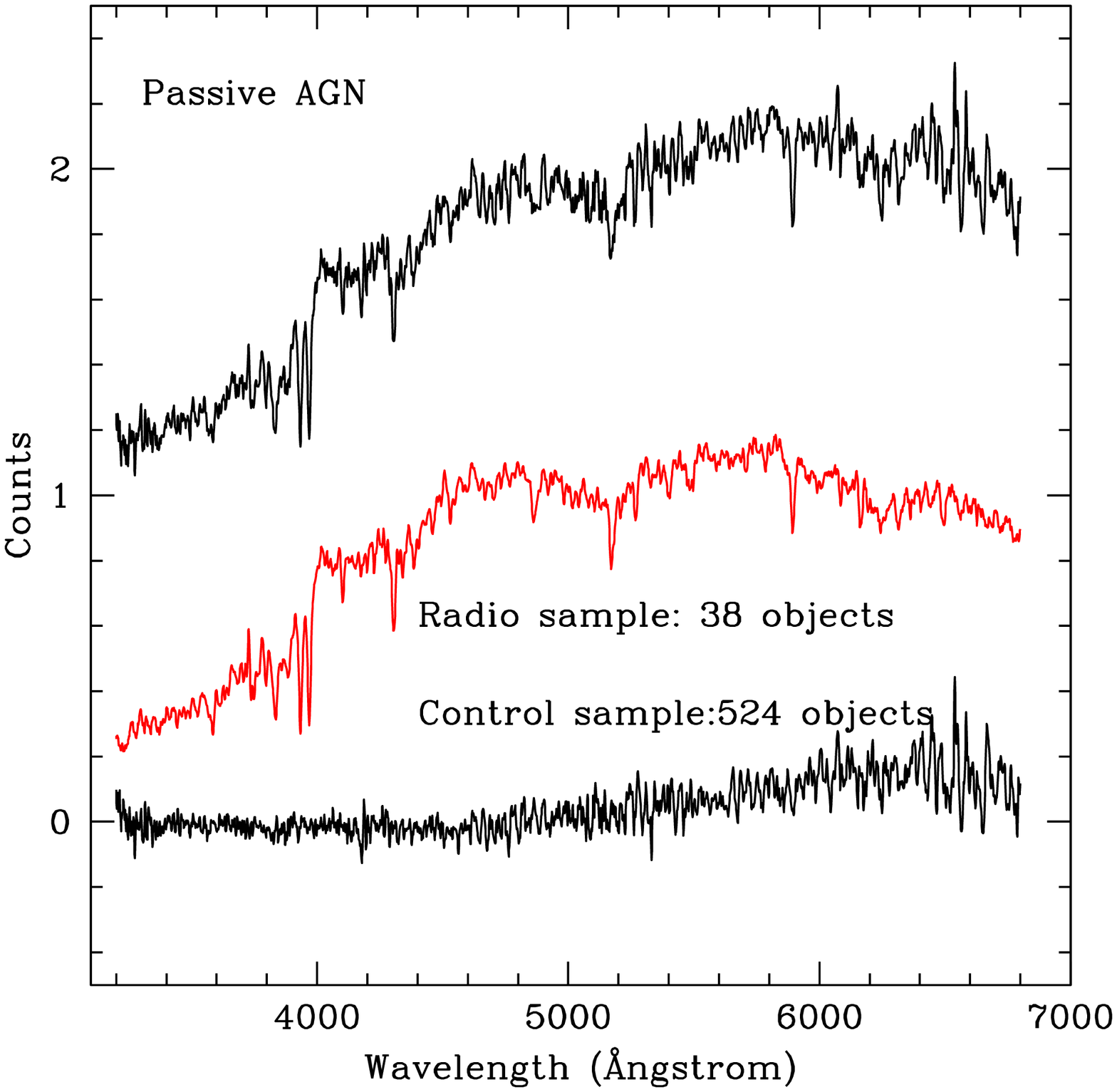}
 \includegraphics[width=\hsize]{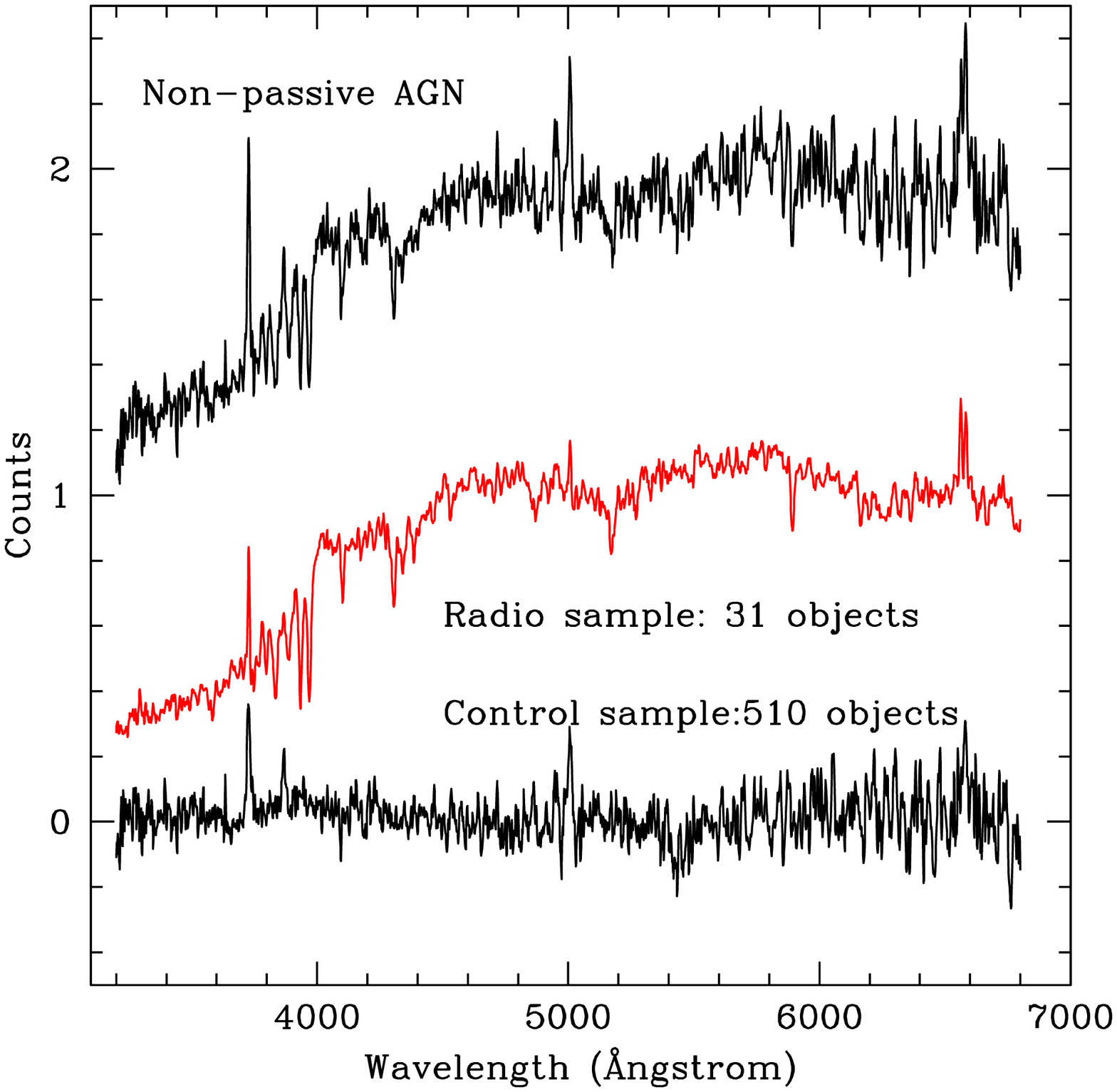}
\caption{Composite spectra for the radio emitting  AGN (in black, upper spectrum) and  the control sample (in red, middle spectrum).
The upper panel refers to passive AGN and the lower panel to non-passive AGN.
The spectrum of radio sources is shifted for clarity by 0.9 in the y-axis. At the bottom of the figures,
we plot the difference between radio emitting objects and control sample (lower spectrum).  }
\label{passivereb3}
\end{figure}

\begin{figure}
   \centering
  \includegraphics[width=\hsize]{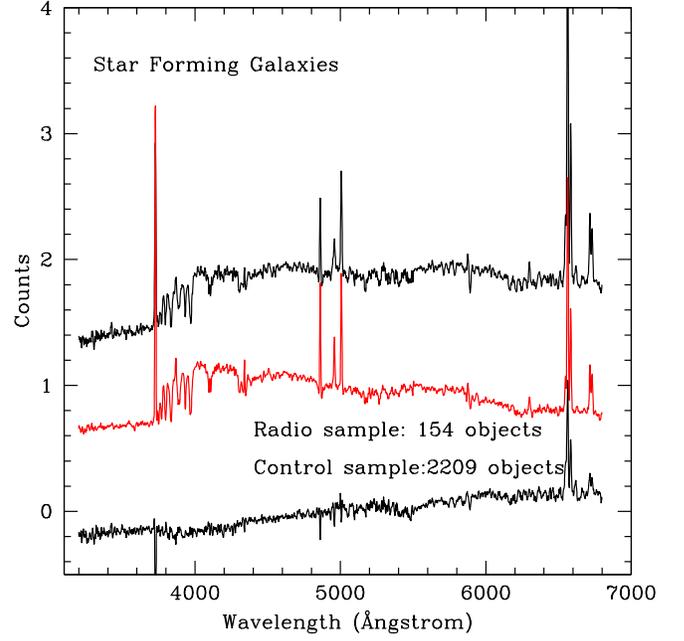} 
 \caption{Same as Figure \ref{passivereb3} but for the radio emitting star forming sample and its control sample.}
\label{blueagnreb3}
\end{figure}

In Figures  \ref{passivereb3} and \ref{blueagnreb3}, we show the composite spectra for the passive, non-passive radio AGN and star forming galaxies  and their control samples. Moreover, we plot the difference between the radio and control sample composite spectra.

For passive AGN, there is no significant  difference between the composite spectra of the radio loud and control sample, implying  that there is no obvious  AGN signature in their optical spectra. There is only an indication for radio spectra to be redder than the control sample.

The spectra of both the non-passive AGN and control sample objects exhibit significant [OII] emission
lines, as well as absorption lines (like H+K Ca and Mg 5175 $\AA$) and relatively weak  [OIII] and $H\alpha $ lines.

In both spectra  H$\beta$ is in absorption only. All these spectral features imply  that 
these galaxies are of relative early type.
All  galaxies from the non-passive AGN and control sample are indeed classified by \cite{Zuccacosmos} 
as spectrophotometric type 1 and 2 (roughly corresponding to Sa-Sb). However, non-passive AGN
tend to be of even  earlier spectrophotometric types than the control sample.  
The radio sample has stronger [OII] equivalent widths  ($15 \AA$ versus $7 \AA$) with respect to 
the control sample.

Considering the star forming galaxies, the difference between the two composite spectra  is negative in [OII]
  and positive in $H\alpha$, implying 
that in radio emitting star forming galaxies a larger amount of line extinction (from dust) is present.
This is confirmed by the fact that the continuum of radio emitting galaxies is redder than the control sample. 
This is related to the differences in colors already found by \cite{Bardelli} for VVDS radio 
galaxies and present also in our VLA-zCOSMOS sample.

Additional  information about the spectral features can be derived from the  equivalent width distribution of the lines. 
We checked this point by examining the distribution of the equivalent width of the [OII] 3727 $\AA$ line 
and the 4000$ \AA$ break \citep[as measured in][]{Mignoli}. There are no significant differences between
the distributions of these quantities for   passive AGN and their control sample. 

Viceversa, the radio detected star forming objects 
have a 4000 $\AA$ break slightly shifted toward higher values than the corresponding control sample.
Given that the stellar mass distribution of the radio detected star forming galaxies and control sample is different  
and that  our data  show a stellar mass- 4000 $\AA$ break relation, this difference is likely due to the different
stellar masses  sampled (see Section 4.2).
 In fact, the differences disappear considering objects with stellar mass $>2\times 10^{10}$ M$_{\odot}$, for which the mass distributions of the radio star forming galaxies and their control sample are identical (see Section 4.2).

\begin{figure}
   \centering
   \includegraphics[width=\hsize]{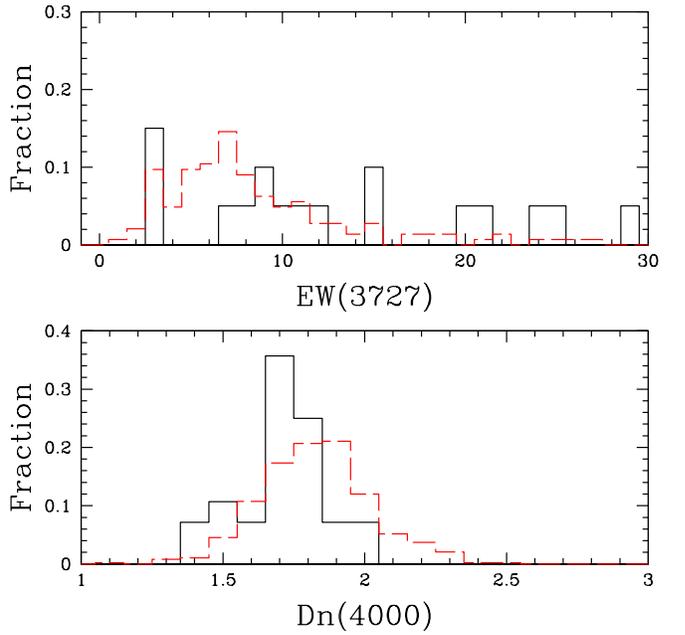}
\caption{ Equivalent width and Dn(4000) distributions for the radio 
and control sample for the non-passive AGN. The dashed red line is for the control sample and
the black solid line is for the radio sample.} 
\label{spectrablueagn}
\end{figure}

More subtle is the difference for the non-passive AGN sample (Figure \ref{spectrablueagn}). In this case
the radio detected objects have higher values for the 4000$ \AA$ break with respect to the control sample and lower 
values for the  equivalent width of the [OII] 3727 $\AA$ line (the probability for the KS test that the two distributions
come  from the same  distribution is $\sim 0.005$).

\begin{figure}
   \centering
   \includegraphics[width=\hsize]{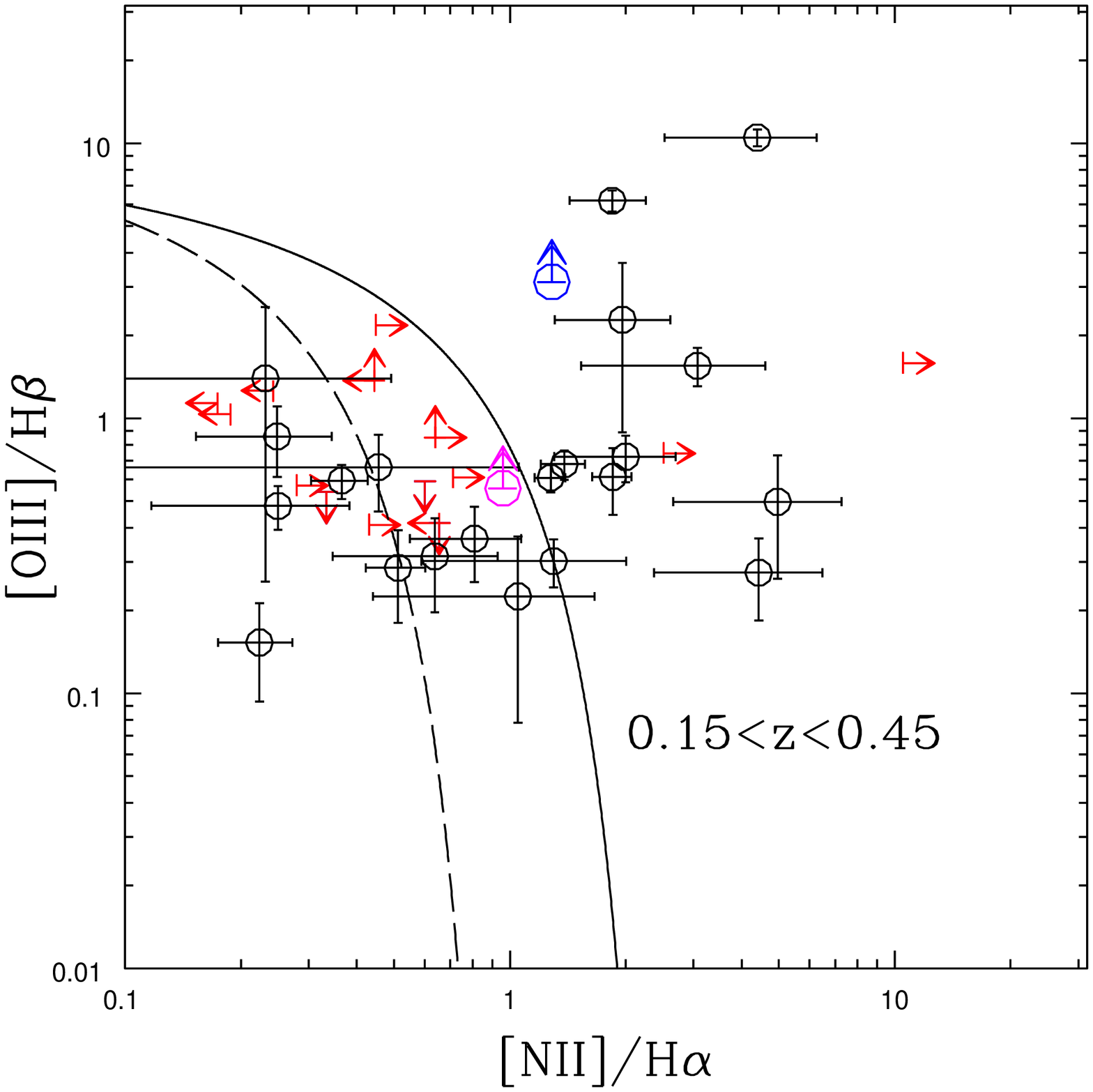}
 \includegraphics[width=\hsize]{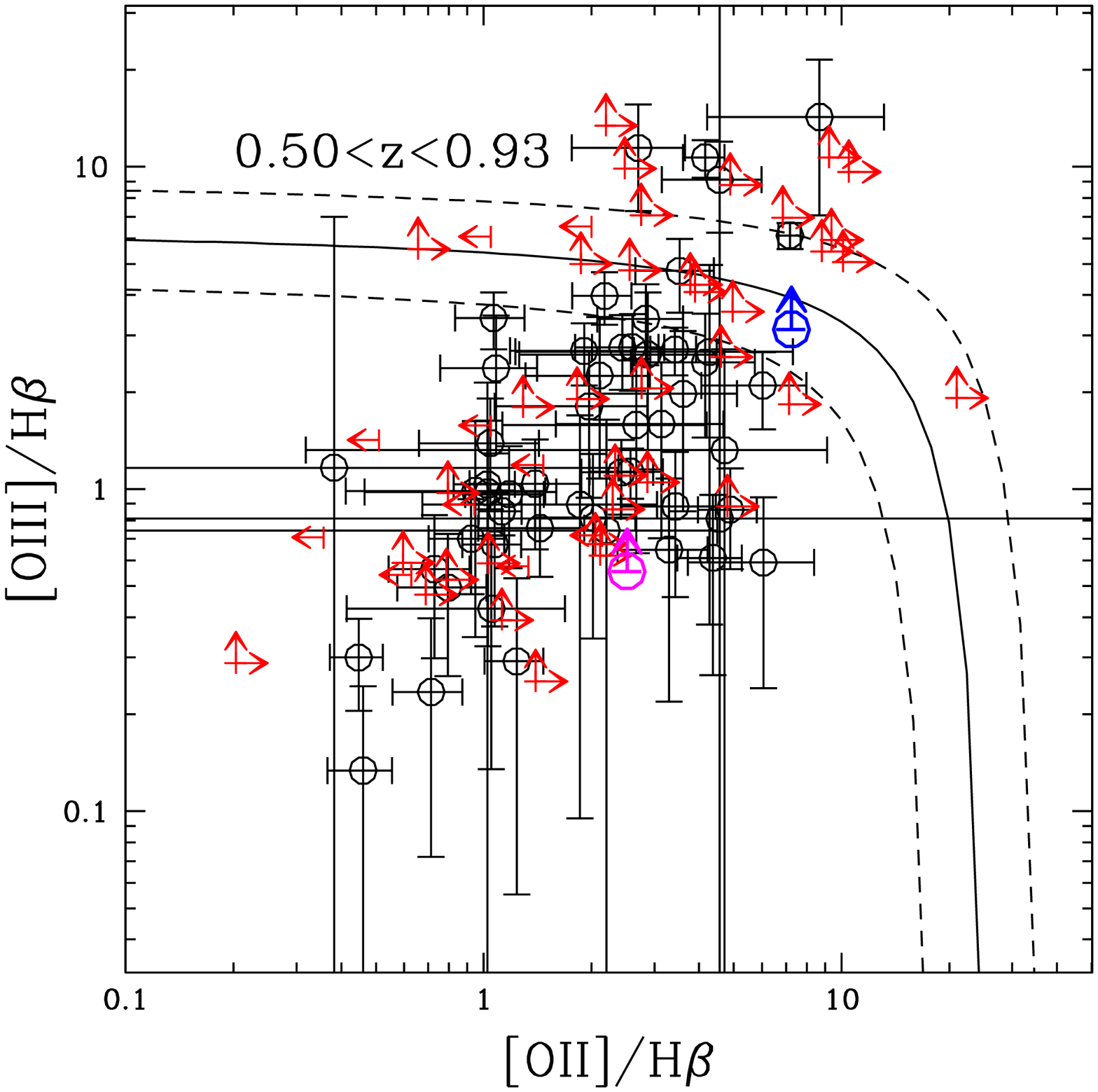}
\caption{Line diagnostics for the control sample of the non-passive AGN sample.  The blue and magenta points correspond to the 
values obtained from the composite spectra of non-passive AGN and control sample respectively. }
\label{diablueagncontrol}
\end{figure}

The last question is whether the control sample of the non-passive AGN shows some sign of AGN activity in their emission lines.
In Figure \ref{diablueagncontrol}, we show the diagnostic diagrams for this sample. The blue and magenta circles correspond to the  values obtained from the composite spectra of non-passive AGN and their control sample, respectively.
Considering the galaxies  with both  all lines detected or  only one line detected in the considered ratio,
$\sim 50 \%$  of this sample consists of  optical AGN or  upper limits consistent with that of AGN. 
Note that this fraction is higher in the red, low redshift diagram.
These fractions are 
broadly consistent  with those of the radio sample, although the number of objects with all detected lines or meaningful upper limits is only 20 for the radio detected 
non-passive AGN sample.

We note that the points corresponding to the composite spectra are not 
in the middle of the distribution of the ratios of single galaxies.
This is  because the composite spectra are obtained also considering objects not plotted in 
Figure \ref{diablueagncontrol} (which are
$\sim 48 \%$  and $27 \% $  of the samples in the blue and red diagram, respectively) because of the lack of detected lines. Therefore, composite spectra are not expected to have the average ratios of the plotted points.


\subsection{Stellar Mass distributions}

 Before studying 
the effect of the different environments on the radio galaxies, it is
necessary to verify the consistency between  the stellar masses of the various samples of radio galaxies
defined in the previous sections with those of their respective control samples.
Given the  mass-density relation \citep{Scodeggio,Bolzonella}, a difference in the stellar mass
distribution may induce a difference in the observed overdensity distributions, which would  not be related to AGN or 
star formation mechanism.

To consider a homogeneous population at all redshifts, we selected galaxies and radio sources 
within an optical  luminosity complete,  volume limited sample at $z<0.9$. To take into account the observed
 optical luminosity evolution \citep{Zuccacosmos},
the absolute magnitude limit was defined to be $M_B=-19.8-z$. 
This limit in absolute magnitude corresponds to the zCOSMOS limit in apparent magnitude ($I<22.5$) at  $z=0.9$.
  In  Figure \ref{massaredcut} we show the observed mass distibution of the passive AGN and  their
control sample. 

For the control sample, since we have a volume limited sample, we plot the stellar mass histogram.
Since the mass completeness limit is changing with redshift (from $\sim 10^9 M_{\odot}$ at $z\sim 0.3$ to 
$\sim 5\  10^{10} M_{\odot}$ at $z\sim 0.9$)\citep[see][]{Pozzetti09},
our observed stellar mass distribution  shows an increasing incompleteness at decreasing stellar mass. 
Assuming that the incompleteness is the same for the control sample and the
radio  sample, only the relative comparison between the control and radio sample is relevant here.
The radio sample, which also has the limit in radio flux in addition to the limit
in apparent magnitude, needs to be corrected for $V_{max}$.

The first result is that all radio emitting objects have $log(Mstar)>10.7$, while the control sample has a 
broader mass distribution.
For this reason, in the following analysis we  limit the control sample to $log(Mstar)>10.7$ .

We note that the ratio of  the stellar mass distribution of radio emitting passive AGN to 
its control sample is rather constant at $\sim 20 \%$.  
This is apparently inconsistent with the results of \cite{Smolcicagn}, for which the ratio of radio AGN to the control 
sample is rapidly increasing with  stellar mass. The difference could be understood considering 
that we study  only a specific class of AGN and  apply a significantly high cut in absolute magnitude
 to create a volume limited sample  at the depth of zCOSMOS.
On the other hand, \cite{Smolcicagn} considered volume limited samples in terms of the radio limit (i.e. fixing a limit 
in the radio luminosity)
 and corrected for $V_{max}$ the optical incompleteness. Their 
apparent magnitude (and consequentely absolute magnitude range) limit is deeper than that of zCOSMOS survey.
 In fact, considering all AGN (passive+blue) and avoiding the absolute magnitude limit, we derive similar ratios as 
\cite{Smolcicagn}.

For non-passive AGN the ratio of the two stellar mass distributions (radio detected and control sample) is $\sim 5 \%$ 
for masses $<10^{11}$ M$_{\odot}$ (see upper panel of Figure \ref{massaredcutbis}), while at higher masses the ratio is $> 25 \%$
We note that in this case we recovered the mass dependence of the AGN activity.

For star forming galaxies, the lower panel of  Figure \ref{massaredcutbis} indicates that the the two stellar masses distributions are the same for masses 
higher  than $2\times 10^{10}$ M$_{\odot}$ and differ for lower masses, consistent with a similar plot in \cite{Bardelli}.
 
 We interpret this trend as implying that the stellar mass distribution
of radio detected star forming galaxies is the convolution of the stellar mass
function of these objects with the star formation rate distribution at fixed stellar mass or 
absolute magnitude \citep[see][]{Bardelli}. 
The mean star formation rate increases with mass, but the distribution has significant tails. 
At  lower stellar massses an increasing 
number of objects with star formation rate in the low tail is below the flux limit of the radio survey.
 More specifically, the lower limit for the star formation rate is set to
$log(\rm{SFR}/Mstar) \sim -10.5$ from Figure \ref{colmstar}, which corresponds for  $log(Mstar)=10.3$ (the "completeness limit") to $\sim 0.6  M_{\odot}$ yr$^{-1}$, similar to the minimum observed star formation rate. For lower masses
the limit is lower than the minimum observed star formation rate.

For the density analysis we limit the control sample to  $log(Mstar)=9.7$ for the star forming galaxies and to 
 $log(Mstar)=10.4$ for non-passive AGN. 
 These limits correspond to the minimum mass detected in the radio samples.

\section{Environmental effects}

\subsection{Method}
To  consistently  define our  radio and control sample
we adopted the limits reported in Table \ref{numtab}. We computed the density
distributions   by considering all objects with $z<0.9$ and brighter than an
"evolving"  absolute magnitde limit $M_B<-19.8-z$, where z is the galaxy redshift. This evolving limit was introduced in order 
to take roughly into account the luminosity evolution of the galaxies as determined  from the luminosity function  estimate by \cite{Zuccacosmos,Zuccavvds} 
and therefore to approximately consider  the same population at all redshifts.
Moreover, in the case of the passive AGN sample, we also applied  a cut in stellar mass
at $log(Mstar)>10.7$, for the non-passive at  $log(Mstar)>10.4$  and for the 
star forming galaxies at  $log(Mstar)>9.7$ (see Section 4.2).

\begin{figure}
   \centering
   \includegraphics[width=\hsize]{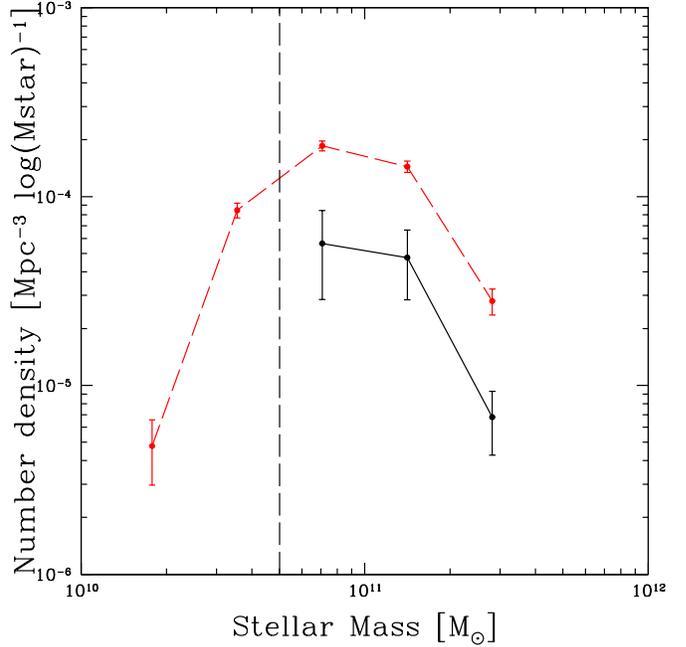}
\caption{Stellar mass distribution of the passive AGN (black continous line) compared with its control sample (red dashed line). The vertical dashed line indicates the
adopted stellar mass cut.}
  \label{massaredcut}
   \end{figure}

\begin{figure}
  \centering
\includegraphics[width=\hsize]{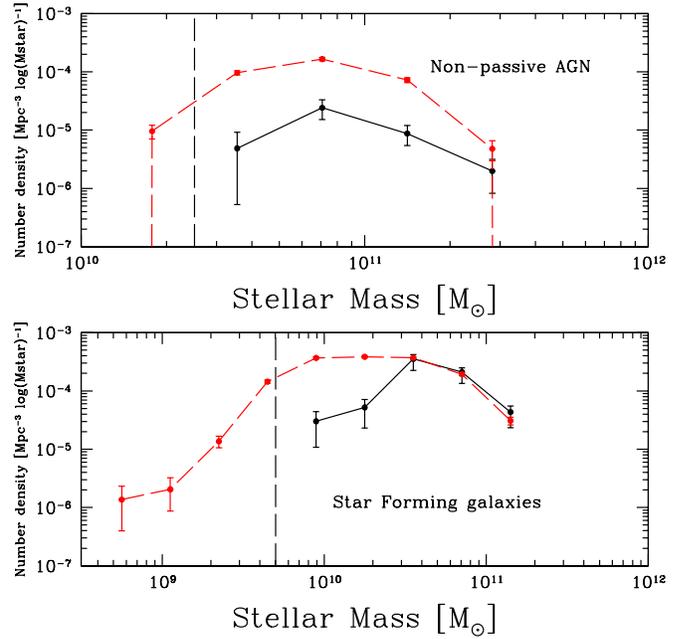}
\caption{Upper panel: Stellar mass distribution of the non-passive AGN (black continous line) compared with its control sample (red dashed line).  Lower Panel: same but for star forming galaxies.}
   \label{massaredcutbis}
   \end{figure}

\subsection{Redshift distribution}

\begin{figure}
   \centering
    \includegraphics[width=\hsize]{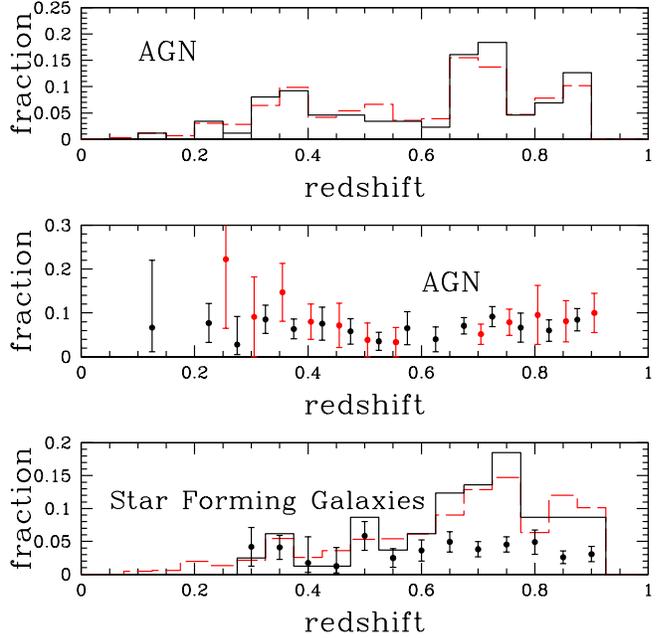}
	\caption{Upper and lower panels: fractional redshift distributions of our samples.  
Red dashed histogram: control sample. Black solid histogram: radio sample. 
Histograms have been normalized to the number of objects.
 Upper Panel: AGN sample (passive+non-passive).
In the middle panel, we show   the ratio between the number
 of radio AGN and the control sample.
Red points represent the fraction of radio loud passive AGN (shifted for clarity by 0.03 in the x axis).
 Lower panel: Star forming galaxies redshift histograms and relative ratios (black points). } 
   \label{zeta}
   \end{figure}

The most basic approach to assess a possible dependence of  radio galaxies  on 
the environment is to compare their redshift 
histogram with that of the control sample. In Figure \ref{zeta} (upper and lower panels) the normalized 
histograms of the radio sources (solid black line) and of the control sample (dashed red line) are shown. 

In this figure, we cut the radio sample at logL$_{Radio}$ (W Hz$^{-1}$)$>23$  to obtain a volume 
limited sample also in terms of radio luminosity.
 Variations in the ratio of the number of radio galaxies to those in  the 
control sample found in correspondence to peaks or valleys of the galaxy distribution
would indicate a different environment distribution between the two populations.
Smoother variations as a function of redshift would imply an evolution of sources.

 In the upper  panel we consider all AGN (non-passive + passive), i.e. those radio sources with
star formation rate excess and, for the control sample, all galaxies with 
$log(SSFR)(optical)< -10.5$.
In the middle panel, we show the ratio of the two histograms in order to visualize better any redshift dependence.
 Errors are presented at the 1 $\sigma$ confidence level. 
Black points correspond to the whole AGN sample, while
 red points refer to passive AGN. For clarity the red points have been shifted
by +0.03 along the x axis.

In the lower panel we show the redshift distribution  of star forming galaxies. 

The number ratios of both AGN and SF radio samples to the respective control samples are rather constant. This
implies that the radio samples follow the same large scale distribution as the 
control samples and therefore no large segregation with the density is present at the largest scales.
For the two classes of AGN taken separately, 
the ratio values are around $0.07$ for the non-passive AGN sample and $0.06$ for the passive AGN sample.

The redshift histograms explore densities which are on the largest scale possible for our redshift survey,
ranging from $\sim 12$ Mpc at $z\sim 0.3$ to  $\sim 21$ Mpc at $\sim 0.5$.
In the following, we  explore smaller scales using the 5$^{th}$ nearest neighbor.

\subsection{The nearest neighbour determined overdensity distribution}

We  used the densities estimated from the distance of the  5$^{th}$ nearest
 neighbour \citep{Kovac} and the volume limited sample as tracer of the density field.
In this case, the tracers are the galaxies with absolute B band magnitude brighter than $-19.5-z$ up
 to redshift 0.7, while for redshifts between 0.7 and 1, the tracers are galaxies with $M_B<-20.5 -z$.  
 This corresponds to overdensities computed on average scales of $\sim 2.4$ Mpc, thus not much larger 
than the typical extent of a cluster.
We also checked the results for the other tracers described in \cite{Kovac}, but the results did not change.
Among the various available weighting schemes,  we used the densities weighted by the stellar 
masses of the galaxies, but  the results did not change either when using the no-weight (i.e. number weighted) estimator.

\begin{figure*}
   \centering
\includegraphics[width=\hsize]{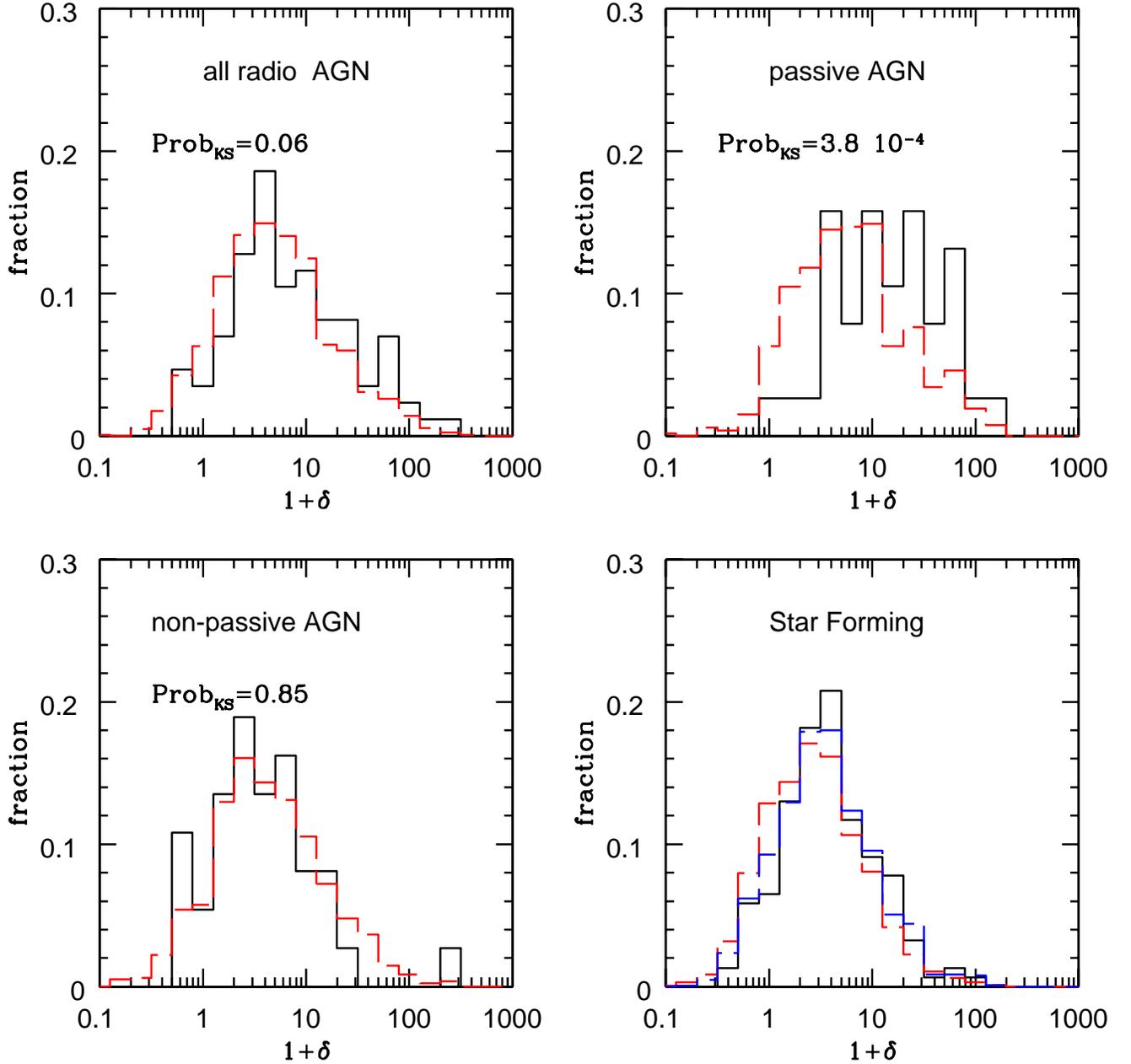}
\caption{Black histograms: overdensity distribution of the radio sources, Red dashed histograms: control samples. Upper left  panel: total AGN sample. Upper right panel: 
Passive AGN sample. Lower left panel: non-passive AGN. Lower right panel: Star forming galaxies. The blue histogram corresponds to the corrected control sample.}
   \label{dens}
   \end{figure*}

In Figure \ref{dens} we compare the overdensity distribution of the three radio samples (black solid histograms) 
with those of their respective control samples.
In the upper left panel,  we present the total AGN sample. We recall that in this case the radio sample 
comprises all radio 
sources
with star formation excess and the control sample comprises all objects with  $log(SSFR)(optical) < -10.5 $.
The two distributions are different at about $2 \sigma$ level on the basis of the K-S test.

\begin{figure}
   \centering
\includegraphics[width=\hsize]{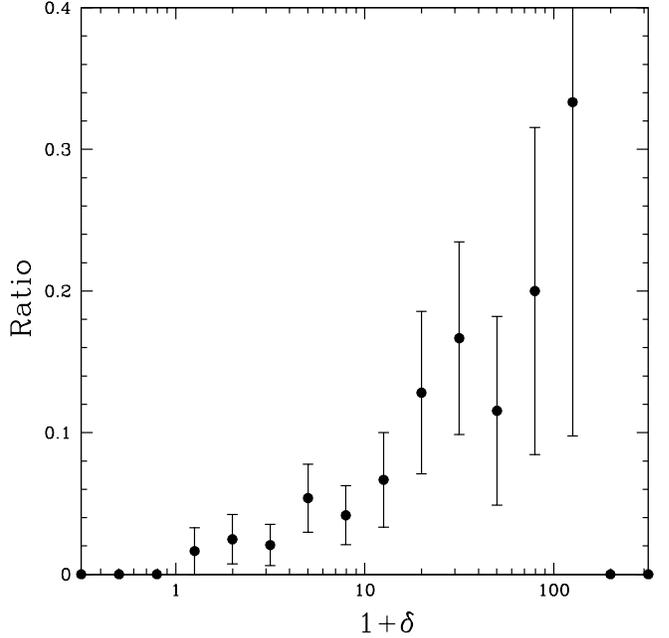}
\caption{Ratio between the radio and control sample of passive AGN as a function of the density.} 
  \label{ratio}
   \end{figure}

 A much stronger statistical signal of difference (as evaluated by the -KS test
probability) is obtained by considering   only the passive AGN (upper right panel).
  Here, we note a lack of radio sources at low  
overdensities ($<3$)  and an excess at high overdensities ($>10$). In fact, the signal present in the whole 
AGN sample is coming entirely from passive AGN. If we select only the non-passive AGN sample,
 the KS test does not find any significant difference (lower left panel). 
In Figure \ref{ratio} we plot the same data but now as ratio of the number of radio emitting passive AGN 
to the control sample. As can be seen, the ratio  steadily increases with the density from 0.02 to 0.2. 

There is the possibility that this behaviour is caused by the fact that the {\it observed} ratio  
of the stellar mass functions of radio to the  control sample shows a strong variation 
between 0.02 (in the bin at lower mass in Figure \ref{massaredcut}) to $\sim 0.2$ (in the highest bin).
 Note that in Figure 
 \ref{massaredcut} we showed the {\it Vmax corrected} mass distributions.
In other words, spurious density differences caused by the stellar mass-density relation could be induced 
both by real differences in stellar mass fucntions between radio and control samples, but also by the fact that
the observed mass distributions are different (although the mass functions are similar) 
merely because of the flux limit.

To check this fact, 
 we recomputed the  control sample overdensity histogram of the upper right panel of Figure 
\ref{dens} by weighting all masses  by the ratio of the observed  mass distribution of the
radio to that of the  control sample. The results changed by a negligible amount.
We checked a possible dependence with the redshift of the difference between the environments of radio and 
control sample. The differences between radio and control samples
remain significant although  the density distributions of both samples
 have higher density tails at lower redshift, consistent with the 
evolution of galaxies and of the overdensities.
We conclude that no redshift evolution is present in the density-radio emission relation.

Another apparently significant difference in environment distribution between radio and control sample is
 coming from the star forming galaxies.
The radio emitting star forming galaxies reside in systematically higher densities than the control sample.
However, we note that also for star forming galaxies the stellar mass functions for the radio  and control sample 
differ (lower panel of Figure \ref{massaredcutbis}).

We repeated the previous procedure to correct the control sample for the stellar mass distributions ratios 
and we find that the significance of the difference disappears.
To do this we computed the   $\chi^2$ probability  of the two distribuions:
we can not use the KS test  because of the binned mass ratios assumed for the correction: the probability 
that the two distributions are the same increases from $\sim 0.05$ to $\sim 0.90$.

 Therefore, we conclude that most of the difference between the overdensity
 distributions of radio and control star forming galaxy samples is caused by the different stellar 
masses distribution  sampled by radio and control samples (higher masses reside in higher densities) 
and not by the physical role of the environment on the radio emission. 

All  results showed in  this paragraph are stable also using the 10$^{th}$ and 
20$^{th}$ nearest neighbor, which correspond to scales of $3.5$ and $5$ Mpc, respectively.

\subsection{Groups of galaxies}

\begin{figure}
   \centering
\includegraphics[width=\hsize]{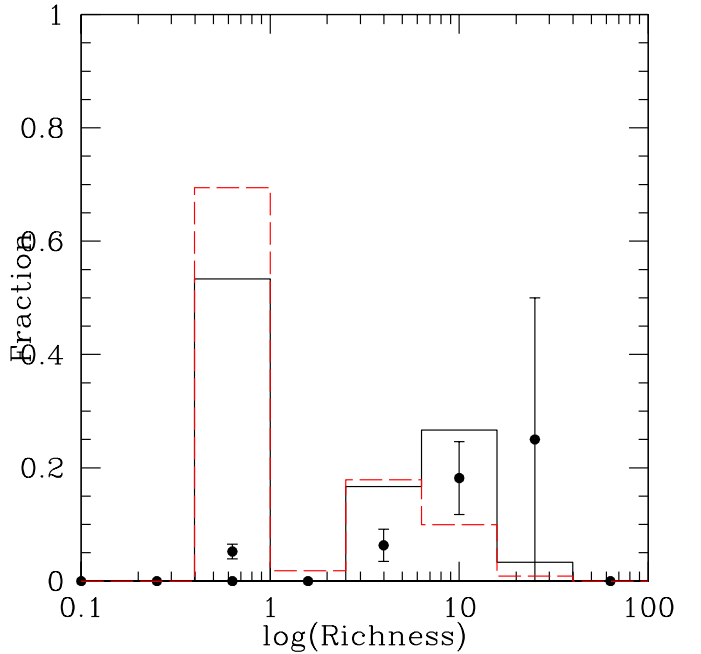}

\caption{
Fractional distribution of richness of the groups where radio passive AGN (solid line) and galaxies of the control sample (red dashed histogram) reside.
Circles with error bars represent the ratio between the numbers of objects of these two samples.} 
  \label{ratiosgrup}
   \end{figure}

In Figure \ref{ratiosgrup} we show the ratio of the number of  radio passive AGN  to the control sample as a function of group richness.
The fraction of passive galaxies which are AGN  in the isolated galaxies sample is $0.052$ (16/307), while the fraction  in groups is $0.104$ (14/135), i.e. the difference is not statistically significant.
It is however worth  noting that 2/3 of the control sample galaxies are isolated and 1/3 are in group, while
the radio passive AGN are nearly equally distributed in the two classes.

 However, if we limit the group sample at richness$>4$
(for which the contamination by false groups is small)  the fraction of radio passive AGN becomes $0.138$ and 
the significance of the difference between the fraction of passive AGN among  isolated  galaxies and in groups
 to $2.15 \sigma$.
 In fact, the overall distributions  of radio and control samples has a significance derived from the K-S test of $8 \%$ to be drawn from the same distribution.
 Therefore, there is a small signal coming from an increase of the fraction of radio objects with  richness.


\section{Properties of the passive AGN}
 
In the previous Section we showed evidence that  passive AGN have an increasing probability to be a radio
source in denser environments. Here, we explore if other properties of single passive AGN also 
change as a function of  local density. 
In practice, we divided the sample of passive AGN  in two bins at a value of  $1+\delta=10$, and compared a number of quantities for the two sub samples.
First of all we computed the two   luminosity functions, 
 We note that in our case it is only possible  to consider  differences in shapes, as the relative normalizations
 are influenced by the volume occupied by the various overdensities within the survey volume, and this is difficult to control.
 For this reason, we compared only the shapes which resulted to be 
indistinguishable.
This is not unexpected because already \cite{LO96} found that the radio and field luminosity functions have similar shape.

\begin{figure}
   \centering
\includegraphics[width=\hsize]{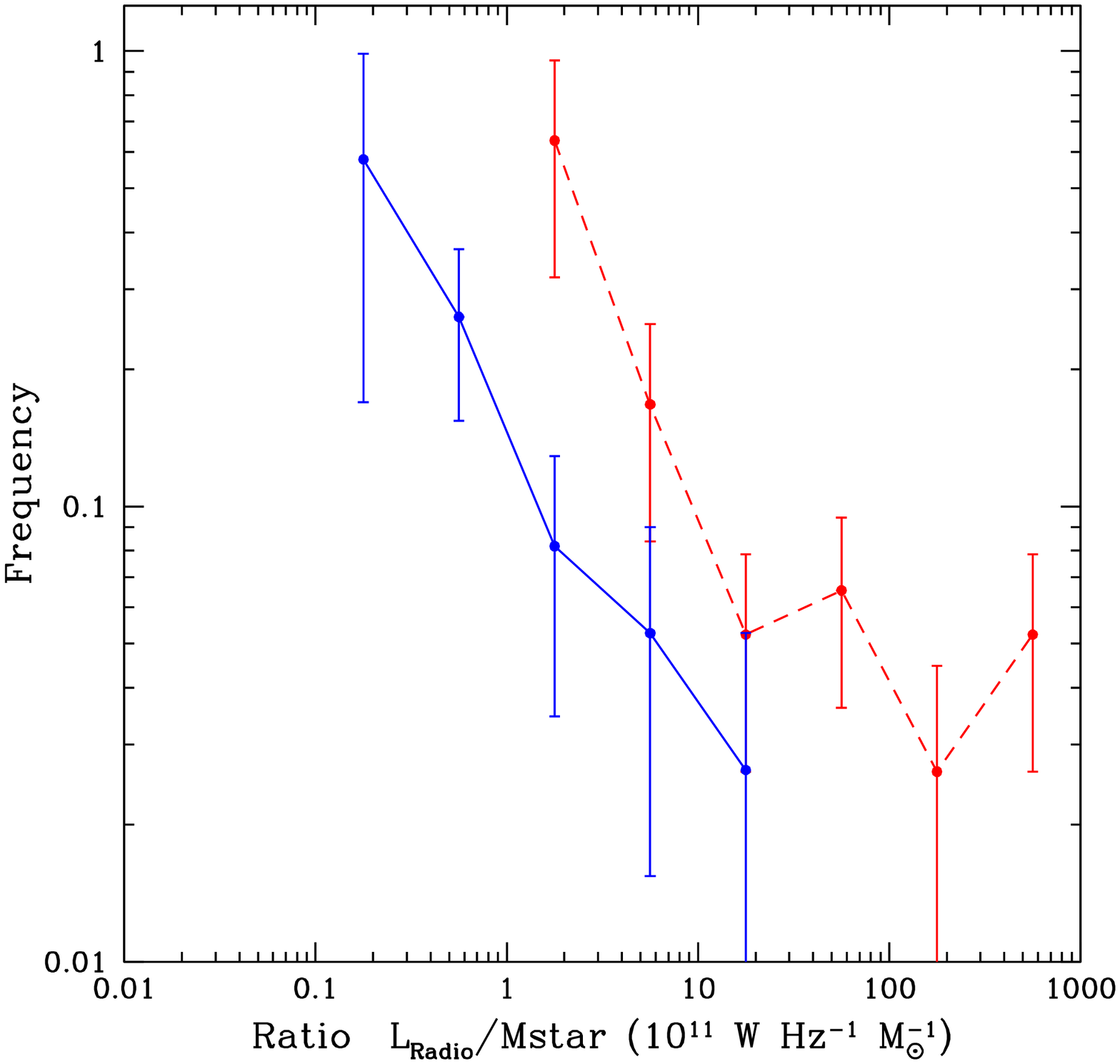}
\includegraphics[width=\hsize]{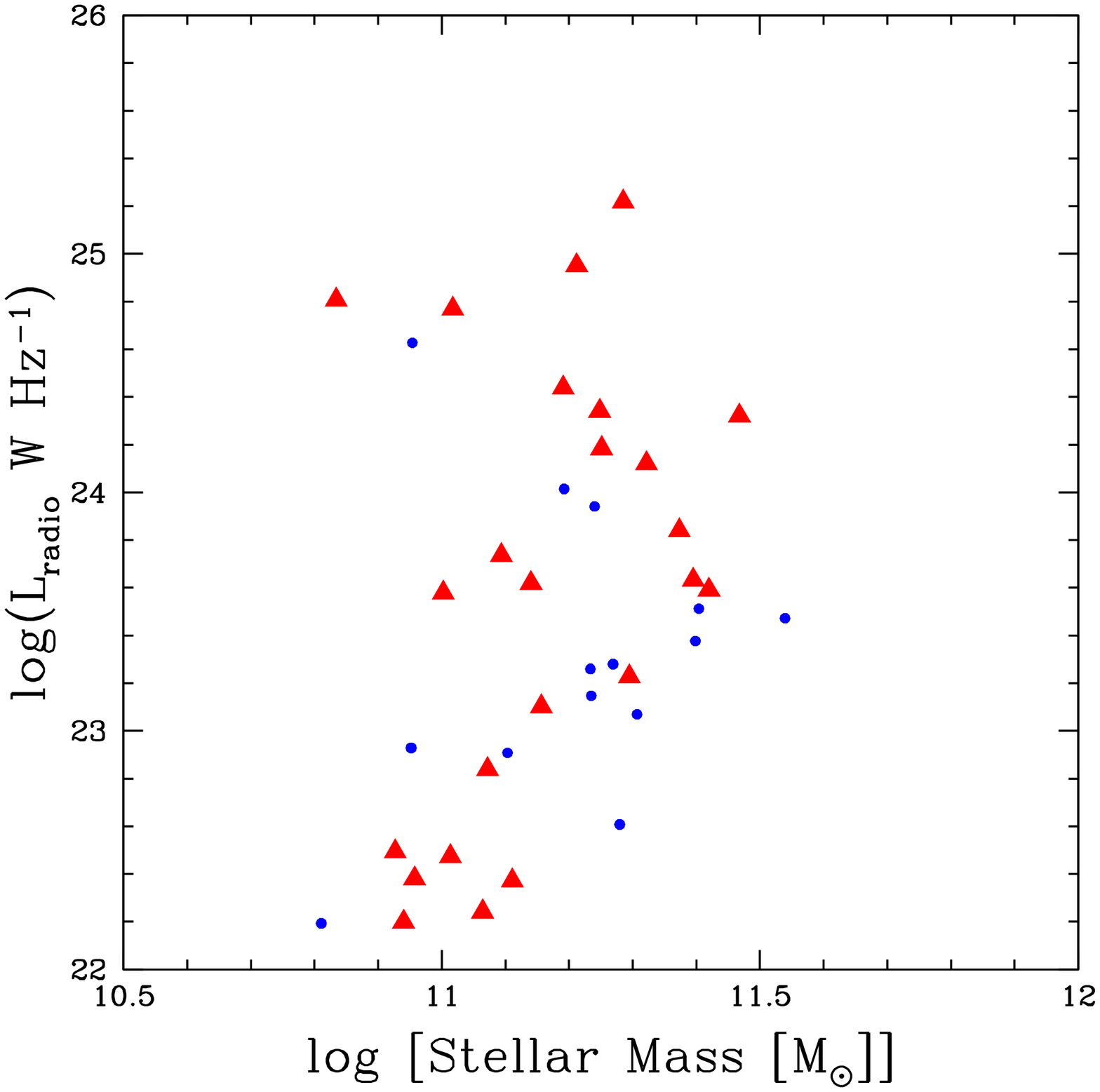}
\caption{Upper Panel: Distribution of L$_{Radio}$/Mstar for passive radio AGN in
two bins of overdensity (red dashed line for high and blue continous line
for low overdensity).
Lower Panel: Radio luminosity versus Stellar Mass for the same samples.
Red triangles represent high densities and blue points the low densities.}
\label{massratio}
  \end{figure}

In Figure \ref{massratio} (upper panel) the radio power-stellar mass ratio is plotted.
Dashed red lines correspond to the high density sample and the solid blue line 
is the low density one. Each ratio has been weigthed by its V$_{max}$ following \cite{Bardelli}.
It appears that, at a given stellar mass, the radio luminosity of the AGN is significantly higher in dense environments.

In the lower panel of  Figure \ref{massratio}, we plot the radio luminosity 
versus the stellar mass. We note that the difference between the two panels is that in the upper one 
the  radio luminosity-stellar mass ratio is corrected by $V_{max}$, while in the lower panel  we plot the observed points.
 It seems that there is a transition luminosity (
at  logL$_{radio}$(W Hz$^{-1}$)= 23.5  ), below which 
the objects are about equally distributed in low and high density regions,
 while above this luminosity $\sim 80 \% $ of radio sources reside
 in high density regions. 
There is also an indication that below the luminosity threshold,
a  correlation between mass and luminosity is present (confirmed by a Spearman
rank correlation test), which disappears at higher luminosities.

\begin{figure}
   \centering
\includegraphics[width=\hsize]{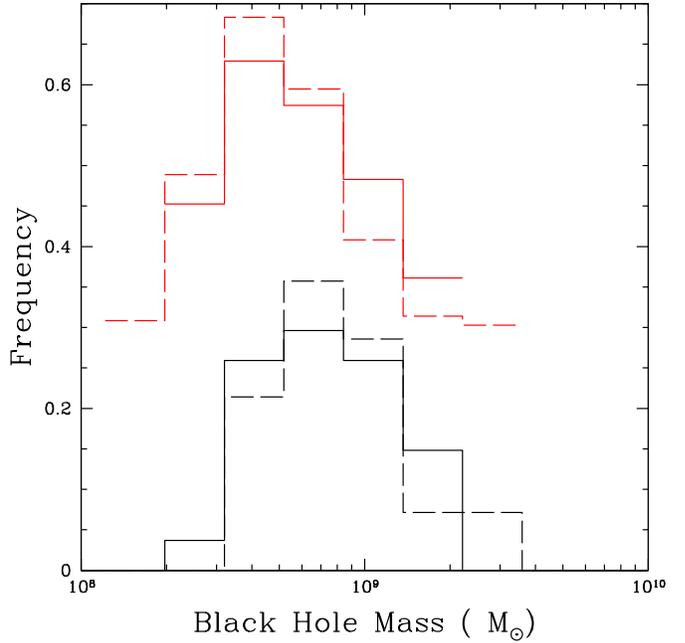}
\caption{Black solid histogram: high density ($1+\delta>10$) black hole masses
distribution for passive AGN. Black dashed histogram: same distribution but for objects with ($1+\delta<10$). In red the distribution of masses for the control sample, shifted for clarity along the vertical axis by 0.3.}
\label{MassBH}
  \end{figure}

In Figure \ref{MassBH}, we show the black hole mass distribution for the passive AGN (black) and the control sample (red). 
We estimated the black hole mass using the prescription of \cite{Marconi}
$$ \log_{10} \left(  { {M_{BH}} \over { M_{\odot} } }\right) = 8.21 +1.13 \left[ 
\log_{10} \left(  { {L_{K}} \over { L_{\odot} }} \right) -10.9 \right] $$
where  $L_{K}$ is the K-band  luminosity of the galaxy. 
We show the results for the local M$_{BH}$-  $L_{K}$  relation,
because \cite{Peng} claim that this relation does not change with redshift.
We also corrected the \cite{Marconi} formula as done by \cite{Smolcicagn}
for the evolution in luminosity of a passive population 
\citep[see][]{Hopkinsbh} but the results did not change. 

There is no significant difference in the mass distribution between 
 high and low density AGN. The only difference is between the high and low 
density control sample (with a KS probability for the two distribution to be equal of  $\sim 0.005$).
 If we plot the radio luminosity versus the black hole mass (not presented here)
we obtain  a  plot similar to the lower panel of Figure \ref{massratio} with the stellar mass substituted by the 
black hole mass. This is reasonable considering the relation between stellar mass and K-band luminosity.

Our general conclusion  is that the main difference between passive AGN in low and high environments is 
 a difference in radio power and that this is due, given the similarity in the black hole masses, to the host galaxies 
or environment structure.

\section{Discussion and conclusion}
 
Studying the environment of radio galaxies requires a division of the radio sources
on the basis of their emission mechanism, because AGN and star forming galaxies may be influenced 
differently by the local density. To do this,
we compared the star formation rate  estimated from the spectral energy distribution  with the one derived from the 
radio luminosity. 
We found a clear correlation between these two quantities for the star forming galaxies:
 we defined as AGN those objects for which
the radio star formation rate is significantly higher than the optical one.
In other words, these objects have a radio luminosity 
significantly higher than that predicted  by their 
star formation. We assume a  stringent cut at which the AGN radio emission is more than
an order of magnitude higher than the star formation one in the same galaxy. In this way we excluded from
our radio-based AGN sample those AGN defined on the basis of their optical emission lines,
such as Seyfert 2 and LINERs, for which the ratio of AGN to star forming radio emission  is lower than the adopted threshold.

Since AGN are hosted primarly by early type optical galaxies and star forming objects 
are basically  spiral galaxies, we expect to recover at least the optical morphology-density relation  
\citep{Dressler}, not related to the radio emission.
Therefore, to investigate whether there is an environmental effect on the radio emission,  we  defined 
control samples to be compared with the AGN and
 star forming radio samples.
To achieve this,  we studied  the
positions of radio sources and  zCOSMOS galaxies in an
infrared color-specific star formation plane. We defined  three populations:
 passive AGN, non-passive AGN and star forming galaxies.
 The passive and non-passive division is reminescent of the low and high-excitacion dichotomy described in \cite{Smolcic09}
 and probably corresponds to two different processes of accretion (hot and cold, respectively). In fact, the two processes
are expected to depend differently on the environment \citep{Hardcastle}.
 The radio luminosity functions of the passive and non-passive AGN are similar in shape but not in luminosity range.
Non-passive AGN stop at logL$_{radio}$(W Hz$^{-1}$)= 23.4, while the passive AGN reach logL$_{radio}$(W Hz$^{-1}$)= 25.2.

 Finally, our radio objects belongs almost all to the FR-I class \citep[as defined by][]{LO96} and the sample lack of 
the broad line AGN class. Therefore we focussed mainly to objects triggered by gas accretion and did not considered 
other mechanism as e.g. mergers.

{\it Passive AGN:}
 This population of AGN is hosted by red galaxies.
Using the morphological classification derived from HST/ACS images through the ZEST algorithm \citep{Scarlata,Tasca}, we find
that $\sim 85 \%$ of both passive AGN and relative control sample 
are classified as early type (i.e. bulge dominated).
 The spectra of these galaxies
with and without radio emission are indistinguishable and the ratio of 
their stellar mass functions is rather constant at $\sim 0.2$ for masses 
higher than $5 \times 10^{10} $ M$_{\odot}$. This value can be interpreted 
as corresponding  to the 
average fraction of time during which the central engine is switched 
on.
 Note that our ratio is derived comparing the mass functions:
if we had used  the observed number of galaxies, we would have found values more similar to those reported
 in Table 4 of \cite{Shabala} ( $\sim 0.5 $ at high masses, $\sim 0.15$ at intermediate masses and $\sim 0.01$ at low masses).

This population exhibits a strong dependence on the environment: the 
fraction of objects with radio emission increases from $\sim 0.02$ for underdense regions ($1+\delta<3$) to $\sim 0.15$
 for overdense regions  ($1+\delta>10$). Similarly, such trend is  also found (but at a low significance level) as a function of  richness of  galaxy groups. 

Studying the radio luminosity-stellar mass plot, we found that for luminosities lower than 
log P(W Hz$^{-1}$)=23.5 there is a correlation between the two quantities, that disappears at higher powers. 
The correlation between stellar mass and radio luminosity 
is not unexpected considering the relation between 
 the accretion on to 
a black hole in the "radio mode" to the black hole mass and hot gas fraction
\citep[see equation 10 of][]{Croton}.
 As stated by \cite{Croton},
the hot gas fraction is approximately constant for galaxies with $v_{vir}>150$ km s$^{-1}$  and therefore
the black hole luminosity scales with the black hole mass, which correlates with the stellar mass.
Interestingly, half of the objects in the low radio 
luminosity regime reside in low densities, while the others are in overdense regions.

The correlation disappears for radio luminosities higher than log L$_{Radio}$(W Hz$^{-1}$)=23.5.
Almost all of these radio luminous objects reside in overdense region. On the other hand,  
no difference is present in the black hole mass between high and low densities
and therefore it seems that in richer environments 
the emission mechanism is more efficient than in low density regions.

Therefore,  we can tentatively say that at low density the emission is 
determined by the host galaxy stellar mass and at high density by the host structure (group/cluster) in which the galaxy resides.
The higher luminosities in denser environments could be caused by a greater fuel supply (due to the central cooling
of the gas)  and/or to the fact that also outside the galaxy there is a significant gas density to confine the radio jet.
On the other hand, we already know that ellipticals at the center of galaxy groups are richer in hot gas than isolated and non-dominant ones 
\citep{Helsdon} and this implies that a mechanism driving gas within 
dominant galaxies is or has been active.

 This can also justify  a longer time
spent by the central engine in the "switched on" status, as suggested by the increasing ratio of passive  AGN to control sample as a function of  density.  
It would be interesting to investigate whether the high density AGN have luminosities that correlate with the 
central cooling flow, following the relation between radio luminosity and accretion mass found by \cite{Mittal}.
Note that our findings are consistent with those of both \cite{Bestenv}, \cite{Mandelbaum}.
 
Finally, the difference in overdensity between radio-loud and control samples remains constant across  the
studied redshift range, implying that no significant evolution in this phenomenon is present.

{\it Non-passive AGN}:
 This is a less well-defined family. In principle, in this class one can find not only AGN but also objects such as post--starburst galaxies.
In  this case, radio emission disappears on a time scale of $10^{8}$ years, which could be longer 
than the disappearing time of the  star formation signatures in optical. However, we expect that these objects are
 rare because of the implied short time scales. In fact, only two of the radio detected post--starbursts 
of the \cite{Verganipsb} sample and one of the upper limits lie within our AGN region.
This means that post-starburst galaxies contaminate our sample by $4 \%$.
The non-passive AGN population occupies a small region of the infrared color-specific star formation 
rate plot, which is approximately  equivalent to the green valley defined in the optical bands. 
These galaxies are objects with  spectrophotometric type earlier than Sa-Sb and applying the 
standard diagnostics based on the emission line ratios we found that $50\% $ are classified as optical AGN and 
the remaining are star forming galaxies. 
The ZEST  morphologies of non-passive AGN are equally divided between
early types and spirals (with a  $\sim 8 \% $ of irregulars), while the control sample has  percentages of $\sim 60 \%$ and $40 \%$, 
respectively.   
The radio emitting objects of this class follow the same environment distribution of the corresponding control sample galaxies and do not 
show the luminosity-density behaviour.

{\it Star Forming galaxies}
The composite spectra of radio emitting star forming galaxies
is redder than the control sample with a smaller [OII]
and higher H$\alpha$ equivalent width.

 Considering that 
the radio detected star forming galaxies are objects where the star formation rate is on average higher than the undetected ones, this implies that  
the dust content is higher. This increase of dust extinction with increasing star formation 
is already  known \citep[see e.g.][]{Pannella} and could justify the relation between the emission lines
and the radio SFR estimators shown in the lower panel of Figure \ref{sfrsfr}.
The ZEST morphological classification shows that the radio detected objects tend 
to be of earlier type than the control sample, as found for the spectrophotometric types.
  
The radio emitting star forming galaxies show an apparent difference in density distribution with respect to the  control sample, but we have shown 
that this is due to the different stellar mass distributions of the two samples.
Correcting the control sample with the fraction of radio emitting objects,
no differences in environment are present between radio and control sample galaxies. 
This is consistent with the conclusion of \cite{Kauffmann2004} that the star formation does not  depend on the density for scales larger that the Megaparsec.

\begin{acknowledgements}
We acknowledge support from an INAF contract PRIN-2007/1.06.10.08 and an ASI grant ASI/COFIS/WP3110 I/026/07/0.The VLA is operated by the National Radio Astronomy Observatory, which is a facility of the National Science Foundation, operated under cooperative agreement by Associated Universities, IIlC.
\end{acknowledgements}

   \end{document}